\documentclass[%
 reprint,
 amsmath,amssymb,
 aps,
 superscriptaddress,
 pra,
]{revtex4-2}

\usepackage[utf8]{inputenc}
\usepackage{graphicx}%
\usepackage{dcolumn}%
\usepackage{bm}%
\usepackage[normalem]{ulem} %
\usepackage[dvipsnames]{xcolor}
\definecolor{myurlcolor}{rgb}{0,0,0.7}
\definecolor{myrefcolor}{rgb}{0.8,0,0}
\usepackage[unicode=true,pdfusetitle, bookmarks=false,bookmarksnumbered=false,
bookmarksopen=false, breaklinks=false,pdfborder={0 0 0},backref=false,
colorlinks=true, linkcolor=myrefcolor,citecolor=myurlcolor,urlcolor=myurlcolor]
{hyperref}
\usepackage{times}
\usepackage[T1]{fontenc}

\usepackage[caption=false]{subfig}
\usepackage{bbm}
\usepackage{comment}

\usepackage{algorithm}
\usepackage[noend]{algpseudocode}

\newcommand{\ie}{i.e.}
\newcommand{\eg}{e.g.}

\newcommand{\ket}[1]{\mathinner{\lvert#1\rangle}}

\newcommand{\Ketbra}[2]{\left|#1\middle>\middle<#2\right|}

\definecolor{blue}{HTML}{4477AA} %
\definecolor{cyan}{HTML}{66CCEE} %
\definecolor{green}{HTML}{228833} %
\definecolor{yellow}{HTML}{CCBB44} %
\definecolor{red}{HTML}{EE6677} %
\definecolor{purple}{HTML}{AA3377} %
\definecolor{grey}{HTML}{BBBBBB} %

\definecolor{orange}{RGB}{255,165,0}
\definecolor{darkgreen}{RGB}{0, 170, 0}
\definecolor{celadon}{RGB}{125, 202, 125}

\usepackage{lipsum}

\begin{document}

 \title{ReQuSim: Faithfully simulating near-term quantum repeaters}

 \author{Julius Wallnöfer}
 \affiliation{Dahlem Center for Complex Quantum Systems, Freie Universit\"at Berlin, Arnimallee 14, 14195 Berlin, Germany}

 \author{Frederik Hahn}
 \affiliation{Dahlem Center for Complex Quantum Systems, Freie Universit\"at Berlin, Arnimallee 14, 14195 Berlin, Germany}
 \affiliation{Electrical Engineering and Computer Science Department, Technische Universit{\"a}t Berlin, 10587 Berlin, Germany}
 
 \author{Fabian Wiesner}
 \affiliation{Dahlem Center for Complex Quantum Systems, Freie Universit\"at Berlin, Arnimallee 14, 14195 Berlin, Germany}
\affiliation{Electrical Engineering and Computer Science Department, Technische Universit{\"a}t Berlin, 10587 Berlin, Germany}
 
 \author{Nathan Walk}
 \affiliation{Dahlem Center for Complex Quantum Systems, Freie Universit\"at Berlin, Arnimallee 14, 14195 Berlin, Germany}
 
 \author{Jens Eisert}
 \affiliation{Dahlem Center for Complex Quantum Systems, Freie Universit\"at Berlin, Arnimallee 14, 14195 Berlin, Germany}
 \affiliation{Helmholtz-Zentrum Berlin für Materialien und Energie, Hahn-Meitner-Platz 1, 14109 Berlin, Germany}
 
 \date{\today}%

 \begin{abstract}
   Quantum repeaters have long been established to be essential for distributing entanglement over long distances. 
   Consequently, their experimental realization constitutes a core challenge of quantum communication. However, there are numerous open questions about implementation details for realistic, near-term experimental setups. 
   In order to assess the performance of realistic repeater protocols, we here present \textit{ReQuSim}, a comprehensive Monte-Carlo based \textit{sim}ulation platform for \textit{qu}antum \textit{re}peaters that faithfully includes loss and models a wide range of imperfections such as memories with time-dependent noise. 
   Our platform allows us to perform an analysis for quantum repeater setups and strategies that go far beyond known analytic results: This refers to being able to both capture more realistic noise models and analyse more complex repeater strategies. 
   We present a number of findings centered around the combination of strategies for improving performance, such as entanglement purification and the use of multiple repeater stations, and demonstrate that there exist complex relationships between them.
   We stress that numerical tools such as ours are essential to model complex quantum communication protocols aimed at contributing to the quantum internet.
 \end{abstract}
 
 \maketitle

\section{Introduction}
\label{sec:introduction}
  Quantum communication constitutes one of the core sub-fields of the quantum technologies. A cornerstone of virtually every large-scale quantum research effort around the world, its most compelling applications include secure communication via quantum key distribution \cite{Roadmap,RevModPhys.74.145,NewQuantumCryptoReview} as well as other multi-party cryptographic primitives \cite{Hillery:1999tb,Murta:2020de} and even functionality for secure, distributed quantum computing \cite{Fitzsimons:2017ge}.
 Indeed, distributed and entangled quantum systems allow us to establish secure encryption keys based on fundamental physical principles.  With respect to real world implementations of key distribution
 over arbitrary distances---in particular arbitrary locations on  earth---it has early on been realized that techniques would be needed that can combat unavoidable losses and errors. In the quantum setting, classical strategies of signal amplification as used
 in classical repeaters are not applicable and so the idea of a \emph{quantum
 repeater} has been devised
\cite{briegel1998quantum,Duer1999repeater,Luong2016,RepeaterLukin}. 
 
 While already to date, impressive implementations
 of direct quantum communication have been 
 achieved \cite{PhysRevLett.121.190502,PhysRevLett.117.190501,Chinese4600km}, there are 
 limits to this approach.
 Only with the help of such quantum repeaters,
can fundamental limits being 
governed by the 
\emph{repeaterless bound}, the so-called PLOB bound \cite{PLOB}
(see Refs.~\cite{Wilde2017, Christandl2017} for a strong converse), be overcome.
Indeed, it is seen that the ultimate limits
in the presence of repeaters are substantially more favourable
\cite{Laurenza,PirandolaRepeaters}. Since the first repeater proposals %
 substantial research effort has been dedicated to experimentally realizing full-scale quantum repeaters.
  These realizations remain a major technological challenge, and therefore, one of the main goals of the quantum communications field is to overcome this significant bottleneck
 \cite{Challenge1,RevModPhys.74.145,NewQuantumCryptoReview}. 
  These technological challenges are not so much conceptual---the basic principles have been 
  known for a long time---but arise from the complicated interplay of the components, which include quantum light and, usually, matter qubits. This central bottleneck is, therefore, primarily one of quantum engineering in the field of quantum optics and light-matter interactions---albeit a persistent and difficult one.

 From a high level perspective, there are indeed a number of open challenges when considering more advanced repeater schemes. For one, there are many design features for variants of repeater protocols that can be modified or combined. Also, it is very 
 difficult to compare advantages between vastly different
 platforms 
 involving trapped ions, NV centres, silicon-based systems or 
 atomic gases. For simple, paradigmatic problems, some settings can be analytically studied, even
 involving some experimentally relevant parameters
 \cite{Luong2016,PRXQuantum.3.010349,ExtendingQuantumLinks, Shchukin2019, Trenyi2020, Kamin2022}. Beyond such paradigmatic settings, 
 an analytical study
 seems to be out of reach. In contrast, several large-scale initiatives which aim at realizing quantum networks
 within the quantum internet---such as the
 \emph{Quantum Internet Alliance} 
 or
the \emph{Quantum Internet Task Force}---have set-up simulation schemes
for simulating high level quantum communication protocols. 
Naturally, they differ in scope and goals as well as the used models and abstractions. Other simulation tools like \emph{NetSquid} \cite{NetSquid}, \emph{QuISP} \cite{satoh2021quisp} and \emph{SeQUeNCe} \cite{sequence} have been used to great effect to study various aspects of quantum networks (see section \ref{sec:related_work} for a more detailed discussion).
 
With \emph{\mbox{ReQuSim}} we present an event based Monte-Carlo simulation platform for realistic quantum repeaters, that is versatile enough to study complex repeater schemes for realistic, small- to intermediate-scale quantum networks and that is first and foremost faithful to a wide range of physically relevant parameters on the hardware level of quantum repeaters. This platform allows us to compare different physical architectures fairly, to explore new regimes, and to identify new, possibly unexpected schemes in the first place. 
In 
\emph{\mbox{ReQuSim}}
we combine 
 \emph{scale} with \emph{realism}. With respect to realism, the platform is detailed enough to, in particular, include arbitrary time-dependent decoherence mechanisms as opposed to a simple waiting-time loss model. This facilitates the modelling of heterogeneous networks incorporating multiple physical realisations for communication channels and network nodes. With regards to scale, our analysis in this work includes schemes of up to 32 repeater links, however, we demonstrate the run-time scaling of up to 1024 repeater links in Appendix \ref{sec:appendix_scaling}. Therefore, we consider our simulation scheme to be useful for any quantum network scenario expected to exist in the short to medium term. 
 
 Analyzing such near-term scenarios with realistic error models is precisely the focus of our method. While there are existing numerical analyses on using realistic devices for quantum repeaters (see,
 e.g., Refs.~\cite{Avis22, NetSquid, Ferreira_da_Silva_2021, sequence}), the sheer variety of potential approaches and the fact that their performance is strongly dependent on the precise circumstances, make further investigation of these concepts indispensable.

 We stress that we can do more than just simulating what will happen in a given scenario with certain resources. Instead, we are able  to simulate a host of scenarios and give actionable advice on what \emph{should} be done with a given set of equipment to maximise performance. 
 This point is critical for establishing whether implementing new methods actually leads to improved results for actual applications.
 We demonstrate such a case in Section \ref{sec:twolink_epp}, where we consider a repeater scheme with a single repeater station that can optionally make use of entanglement purification as an additional tool. While this certainly increases the complexity of the scheme, there are clear circumstances where it outperforms a naive scheme using only entanglement swapping. However, discarding qubits in memory after a certain cutoff-time, is an alternative method that can be used to reduce the error rates at the cost of throughput. A proper optimisation over cutoff times (a relatively simple software adjustment instead of requiring new operations) can see the relative advantage shrink or vanish. 
 In this case, the overall performance would not be meaningfully improved by the expenditure of additional resources on employing entanglement purification methods. This kind of detailed benchmarking is essential to properly evaluate under which circumstances a particular building block should be utilised.

 With \emph{\mbox{ReQuSim}} we can provide guidance in a meaningful strategy analysis. 
 Using multiple repeater stations instead of one can 
 extend the reachable distances, but will also subject the process to additional noise from storing more qubits in memory and performing more operations. In Section \ref{sec:multilink}, we explore this trade-off and show that the number of repeater stations needs to be carefully optimized if their quality is limited, even if their quantity is not.
 Furthermore, we demonstrate that improving the quality of those repeater station resources results in a disproportionate improvement in achievable rates, as it allows us to modify the protocol to use a larger number of repeater stations, which would otherwise be detrimental. In a similar vein, in Section \ref{sec:many_params} we demonstrate how our approach can be used to give insight about necessary hardware parameters in a setting with a customized noise model.

 Considering the points above, it is clear that there is no shortage of potential options to improve performance. As the examples shown in this work compellingly demonstrate, there are highly nontrivial relationships between \emph{combinations} of these strategies. 
 Optimising repeater protocols for a variety of realistic devices therefore requires a deeper understanding of repeater protocols and is not simply a matter of tweaking a few parameters.

 \begin{figure*}
    \centering
\includegraphics[width=\linewidth]{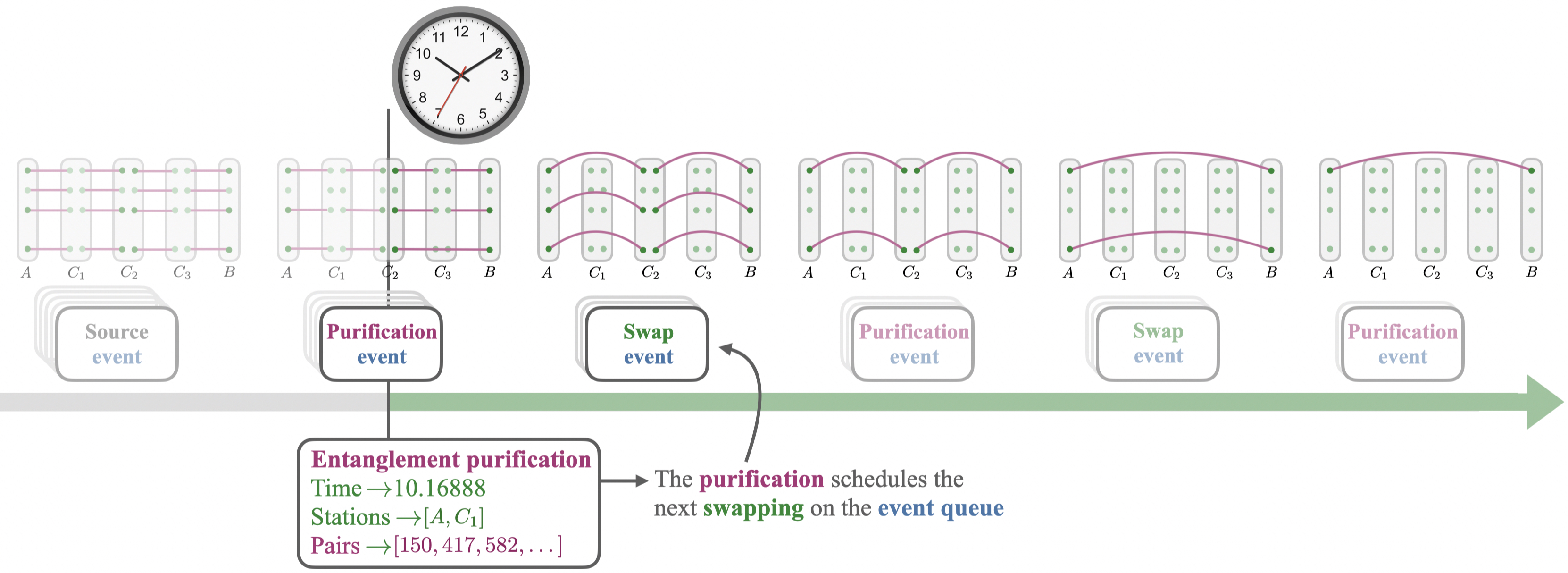}
    \caption{\label{fig:Simulation_Illustration} 
    An entanglement-swapping-based repeater protocol in the \emph{\mbox{ReQuSim}} simulation framework. 
    Running the protocol schedules first events on the \textcolor{blue}{event queue}. 
    The events on the queue are \textcolor{grey}{successively} \textcolor{green}{resolved} and may schedule further events. 
    An example of this is highlighted above for an \textcolor{purple}{entanglement purification event} scheduling an \textcolor{green}{entanglement swapping event}.
    }
\end{figure*}

\begin{figure}
    
    \begin{algorithm}[H]
        \caption{Run multi-memory repeater}
        \begin{algorithmic}
            \State $\mathsf{scenario} \gets$ 
            Setup
            including \eg,~position of 
            stations and 
            sources
            \State $\mathsf{iter} \gets$ desired 
            number %
            of long distance 
            entangled
            connections
            \State $\mathsf{params} \gets$ parameters of the model as described in 
            Section \ref{sec:noise_model}
            \State $\mathsf{protocol} \gets$ 
            description of the repeater protocol to perform

            \Function{run}{$\mathsf{scenario}, \mathsf{iter}, \mathsf{params}, \mathsf{protocol}$}
            \State \textsc{initialize} 
            setup %
            according to $\mathsf{scenario}$ \& $\mathsf{params}$
            \State $l \gets$ number of successfully established links starting at $0$
            \While{$l$ < $\mathsf{iter}$} %
                \State 
                $\mathsf{protocol}$.\textsc{Check} for new events to be scheduled
                \If{long distance connection has been established}
                    \State \textsc{Update} collected data.
                    \State $l \gets l+1$
                \EndIf
                \State \textsc{Resolve} next event in event queue
            \EndWhile
            \EndFunction

            \Function{Check}{} 
            (example for a single station)
                \State $n_\mathrm{mem} \gets$ number of quantum memories per link at station
                \State $n_{l,r} \gets$ number of established pairs %
                left/right of 
                station
                \State $s_{l,r} \gets$ number of 
                events creating pairs 
                left/right of 
                station 
                (\ie~accounting for busy memories with no established pair)
                \If{$n_{l} + s_l < n_\mathrm{mem}$}
                    \State \textsc{Schedule} source events %
                    drawing
                    from time distribution 
                \EndIf
                \If{$n_r + s_r < n_\mathrm{mem}$}
                    \State \textsc{Schedule} source events
                    drawing
                    from time distribution 
                \EndIf
                \If{$n_l > 0$ and $n_r > 0$}
                    \State \textsc{Schedule} entanglement swapping event 
                \EndIf
            \EndFunction
            
        \end{algorithmic}
    \end{algorithm}
    \caption{Pseudocode for formulating a protocol and running the simulation. This example describes a protocol for a single repeater station that has access to multiple quantum memories to simultaneously attempt to establish an entangled connection for both repeater links and immediately performs entanglement swapping once successful qubits for both directions are in memory.
    The \textit{events} are standardized ways to interact 
    with the current state of the simulation. They have a method that specifies the associated quantum operation that is performed on the quantum states when they are resolved, e.g., a Bell state measurement for the entanglement swapping event.
    }
    \label{fig:pseudocode}
\end{figure}

\section{ReQuSim: The simulation framework}
 \emph{\mbox{ReQuSim}} \cite{requsim} is a simulation framework we have developed specifically for simulating quantum repeaters. It is available as an open source Python package from the 
 \emph{Python Package Index}.
 Simulating quantum repeaters
 is a very attractive option, as the complexity of analytical expressions is increasing rapidly if one deviates from the known standard cases. A central challenge in that regard is that everything that happens anywhere in the system can potentially effect parts on the opposite end of the repeater chain, \eg,~the time a qubit needs to wait in memory. Working with averages calculated for parts of the system does not necessarily give the full picture if there are non-linearities such as the dephasing in quantum memories. By contrast, Monte-Carlo methods are well suited to deal with systems with many interconnected probabilistic components.
 The simulation approach further allows for a modular design that makes switching to different error models and introducing asymmetries straightforward.

 In the following, we describe the basic working principles of the simulation.
 First, the various parts needed for the given scenario are initialized, e.g., where the end stations and repeater stations are located. This also includes setting up the noise models that are specific to certain devices, such as the dephasing time for quantum memories.
 
The core of the simulation relies on an event system the repeater protocol can interact with. Changes to the state of the simulation are done via events that represent operations being performed. These are scheduled in an event queue and later resolved at the appropriate time. Naturally, an event needs to have the information about when to resolve and which are the involved quantum states and repeater stations. For example, an entanglement swapping event would follow the data structure $(\text{type, time, pairs, station})$. 
 What exactly occurs when an event resolves depends on the type of the event, \eg,~an entanglement purification event will simulate the required quantum operations and measurements for an entanglement purification protocol on the involved quantum states.

 The other moving part of the system is the repeater protocol, \ie~the high-level strategy, the simulation should follow, because this defines which events should be scheduled and when.  Naturally, the chosen repeater strategy decides which events exactly are used, \eg,~a repeater protocol without entanglement purification may decide to schedule an entanglement swapping event at a station, immediately after entangled pairs in both directions have been established. Alternatively, in Fig.~\ref{fig:Simulation_Illustration} the high-level strategy of a repeater protocol following a layered approach for entanglement swapping is illustrated. 

 The way the simulation moves forward revolves around a constant interaction of the protocol and the event system. That means after an event is resolved, the protocol immediately checks whether any new events need to be scheduled, \ie~the protocol will continuously monitor the changes performed by the last event. An abstract representation of how this core simulation loop and the decision making process of a simple repeater protocol works is described in Fig.~\ref{fig:pseudocode}. 

\begin{figure}
    \centering
    \includegraphics[width=\linewidth, trim={15cm 0 15cm 0}, clip]{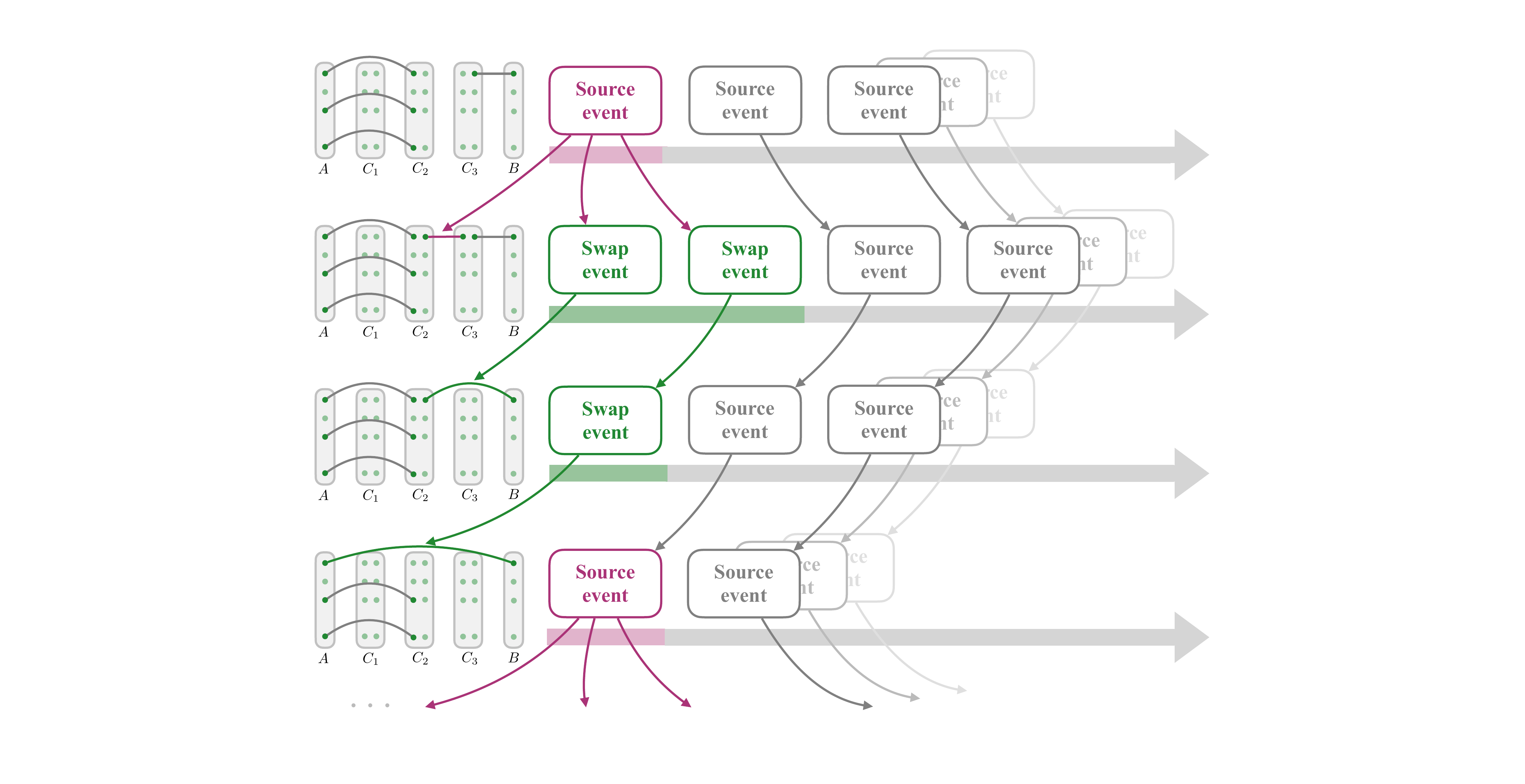}
    \caption{\label{fig:time_steps} Illustration of a step by step update of the state of the simulation and the event queue. The source event (representing the successful establishment of an entangled pair between neighboring repeater stations) triggers the protocol to schedule two new entanglement swapping events, inserting them at the front of the event queue. Resolving the entanglement swapping events establishes a long-distance pair between the end stations. After each step, \ie, after every event that is resolved, the protocol again has the chance to react to the new situation.}
 \end{figure}

 While protocols are usually formulated from a high-level point of view similar to the situation depicted in Fig.~\ref{fig:Simulation_Illustration}, due to the probabilistic nature at multiple parts of the simulation, \eg, the initial distribution process, the actual simulation process needs to be able to act and react at a more detailed level. Consider as an example the situation in Fig.~\ref{fig:time_steps}: While multiple pairs between station $A$ and $C_2$ have been established, the protocol is currently waiting for progress on the right hand side before it can proceed further. After a source event confirms the successful generation of an entangled pair between stations $C_2$ and $C_3$, the protocol reacts by scheduling entanglement swapping events at stations $C_2$ and $C_3$, since these are now finally possible. Since these can be performed immediately, they are inserted at the front of the event queue, pushing back the other events that happen at a later point in simulated time. Then, the entanglement swapping events are resolved one by one, and after each one the protocol checks whether anything new needs to be done---in this case no new events need to be scheduled. Finally, the event system goes back to resolving the previously scheduled events in the event queue.

  The output of the simulation is a sample of the time and state distributions that is produced by the repeater setup. This means, one gets a list of states (if distributing entangled pairs) or error probabilities (if measuring qubits arriving at the end stations as soon as possible, as one would do for quantum key distribution) and the times when they and all necessary classical information were present at the end stations. 
  For the purposes of this work we focus on key rates for quantum key distribution as a familiar way to assign meaning to both the speed and quality of the entanglement distribution process. However, the simulation is not limited to this application and the obtained information can easily be used to calculate other figures of merit like the raw rate of distributed pairs or the fraction of pairs above a certain fidelity threshold. For the key rates in particular, we let the simulation run until a large sample (usually $10^5$) has been obtained and then we use the sample mean of the error rates to estimate the asymptotic key rate.

 Before moving on to the particular scenarios discussed in this work, we comment on a few design decisions of our approach:
 This system centered around \emph{events} is very flexible when it comes to including noise models as the events can simply be modified depending on the conditions, \eg,~the repeater station from the example above may have a parameter that describes the quality of Bell state measurements available at this particular station. \emph{\mbox{ReQuSim}} supports arbitrary completely positive, trace preserving maps to describe physically 
 meaningful noise processes.
 
 Two key challenges that are addressed by our framework are the probabilistic nature of distributing entangled states between neighbors 
 and the time-dependent noise that acts on qubits stored in quantum memories. 
 Generally, the probability that a photon gets lost in transmission is very high, so it is certainly not efficient to keep track of every single photon that is sent individually. Instead, we draw from a probability distribution for how many trials are needed until the next successful attempt happens (this is simply a geometric distribution if the probability of success in one trial stays constant). This means we use a combination of two methods to handle probabilistic aspects of the simulation -  a Monte Carlo approach for loss and a the density matrix formalism for other types of noise.
 We handle
 the time dependent nature of the noise in quantum memories not by continuously updating the quantum state \eg,~in fixed time steps, but instead by doing so only when it becomes relevant, \eg,~when performing an operation involving this particular entangled pair. 
 For this we simply need to track how much time has passed since the last time this update was performed. Both of these approaches are essential for allowing the simulation to run in a reasonable time frame.
 
 At various points in a quantum repeater protocol, classical information needs to be exchanged. While, again, we do not simulate each individual message and are not concerned with the exact content of the classical communication, it is nonetheless essential to take the timing of classical information into account. This we achieve in our simulation by adding an appropriate delay to when an event will be able to be resolved or by blocking any additional operations being performed on a certain part of the simulation until a point in time when the necessary classical information is able to reach the involved parties.

\section{Model}
\label{sec:noise_model}

We consider two distant parties that want to share a secret key via an entangled swapping based repeater protocol with one or multiple repeater stations between them. The repeater station have the capability to locally perform quantum operations such as entanglement swapping and are equipped with quantum memories. Some of the stations also control an entangled pair source that they can use to establish entangled links between neighboring stations.

In any realistic setting, one inevitably needs to account for imperfections in multiple parts of the system, which make employing a quantum repeater protocol necessary in the first place. 
In the following paragraphs we describe a number of sources of imperfection and how we modeled them.
Additional comments can be found in Appendix~\ref{sec:detailed_model}.

\paragraph{Arrival probability.}
Establishing links between neighbouring stations is an elementary operation for the protocols we consider. 
However, creating entangled links between stations is not always successful and usually needs multiple tries. 

There are sources of loss that occur systematically, regardless of the precise layout of the repeater stations. These could for example include the probability of an entangled pair being generated in the first place (preparation efficiency), the wave length conversion efficiency, the probability that photons are successfully coupled into the optical fiber, the efficiency of the detectors or the probability qubits are loaded into quantum memories successfully (memory efficiency). We summarize these in an abstract success probability $P_\mathrm{link}$, which represents  the probability that a pair can be established not taking into account distance-based losses.

Another central source of loss in the system is the distance dependent loss of qubits during transmission.
We describe this by the channel efficiency $\eta_\mathrm{ch}$. For the purposes of this work we always consider photons being sent through optical fibers, but for other repeater setups (such as quantum repeaters using satellites as repeater stations), this part would need to be modified. We define
\begin{equation}
    \eta_\mathrm{ch}(L) := e^{-\frac{L}{L_\mathrm{att}}},
\end{equation}
where $L$ denotes the length the photon has to travel and $L_\mathrm{att} = 22 \text{ km}$ denotes the attenuation length of optical fibers at Telecom wavelengths.
Therefore, the total probability $\eta$ that a pair is established between two neighboring repeater stations separated by a distance $L$ in one trial is given by
\begin{equation}
    \eta = P_\mathrm{link} \times \eta_\mathrm{ch}(L) .
\end{equation}

\paragraph{Initial fidelity.}
The initially created and distributed Bell pairs may also be imperfect. This could stem from an imperfect generation procedure of the entangled pair sources or from other systematic errors in the handling of the qubits of the initial pairs. 

We assign an initial fidelity $F_\mathrm{init}$, which represents all these imperfections except the effect of dark counts discussed below, which is loss-dependent.
The initial state  given by depolarizing noise acting on the desired Bell state vector $\ket{\Phi^+}$ as
\begin{eqnarray}
    \rho_\mathrm{init} &=& F_\mathrm{init} 
    \Ketbra{\Phi^+}{\Phi^+}\\ 
    &+ &
                         \frac{1-F_\mathrm{init}}{3}
                         (
                         \Ketbra{\Phi^-}{\Phi^-} + 
                         \Ketbra{\Psi^+}{\Psi^+} + 
                         \Ketbra{\Psi^-}{\Psi^-} 
                         ).\nonumber
\end{eqnarray}

\paragraph{Dark counts.}
Dark counts are an imperfection in detectors and manifest themselves by \textit{clicks} when no actual signal has arrived. Usually, this is measured in dark counts per second, but for our purposes we are interested in the probability, that a dark count will occur in a detection window, when we would potentially expect a signal. We denote this probability by $p_d$, \eg,~for a detector with a dark count rate of 1 Hz and detection windows of 1 $\mu\text{s}$ we get $p_d=10^{-6}$. 

We can model dark counts by replacing the erroneously expected qubit with a fully mixed state. In order to obtain the effective density matrix we need not only consider the probability that a dark count occurred, but we must set this into relation with the probability of a true click occurring in the first place, \ie~when success rates are very low, nearly every click is caused by a dark count.

The chance for a detector to click is given by
\begin{equation}
    \eta_\mathrm{eff} = 1 - (1-\eta)(1-p_d)^2,
\end{equation}
and the probability that the click indicates a real event is then
\begin{equation}
    \alpha(\eta) = \frac{\eta(1-p_d)}{\eta_\mathrm{eff}} .
\end{equation}
Therefore, the state $\rho$ that represents the output state of a successful attempt is affected by
\begin{equation}
    \alpha(\eta) \rho + \frac{1-\alpha(\eta)}{2}(\mathrm{tr}_i \rho) \otimes \mathbbm{1}^{(i)},
\end{equation}
where $\mathrm{tr}_i$ denotes the partial trace over the $i$-th subsystem and $\mathbbm{1}$ is the identity operator.

\paragraph{Memory noise.}
Since the usage of quantum memories is of central importance to this type of quantum repeater, time-dependent decoherence for qubits stored in them is another source of imperfection. We assume that the noise caused by this is predominantly in one direction, and described by a dephasing noise channel
\begin{equation}
    \mathcal{E}_z^{(i)}(t) \rho = (1-\lambda(t)) \rho + \lambda(t) Z^{(i)} \rho Z^{(i)}
\end{equation}
with the index $i$ marking the qubit that is stored in memory and 
with
\begin{equation}
    \lambda(t) = \frac{1 - e^{-t/T_\mathrm{dp}}}{2},
\end{equation}
where the dephasing time $T_\mathrm{dp}$ is a parameter specific to the quantum memory in question. Note that other imperfections of the memories, \eg,~related to read-in and read-out of qubits can be included in other parameters such as $P_\mathrm{link}$ and imperfections in the Bell measurements.

\paragraph{Imperfect Bell measurements.}
Entanglement swapping is the main quantum operation that needs to be performed once the repeater links have been established. Naturally, the necessary Bell state measurement will be subject to imperfections, which we model as two-qubit depolarizing noise (acting on the qubits being measured) followed by the perfect Bell measurement.
The two-qubit depolarizing noise acting on qubits $i$ and $j$ is given by
\begin{equation}
    \mathcal{E}_w^{(i,j)}(\lambda_\mathrm{BSM}) \rho = \lambda_\mathrm{BSM} \rho + \frac{1-\lambda_\mathrm{BSM}}{4} \left(\mathrm{tr}_{i,j} \rho\right) \otimes \mathbbm{1}^{(i,j)},
\end{equation}
where $\mathrm{tr}_{i,j}$ denotes the partial trace over the $i$-th and $j$-th subsystem and $\lambda_\mathrm{BSM}$ is the Bell state measurement ideality parameter ($1$ corresponds to perfect operation), which is a property of the repeater station where the Bell state measurement is performed.

Another common imperfection is that the measurement may only work with a certain probability, which not only reduces the reliability of the setup, but can even change which swapping strategy should be used \cite{Shchukin2022}. However, for the purposes of this work, we consider the Bell state measurements to always be successful.

\paragraph{Distribution times.}
In the scenarios considered in this work, the entangled pair sources are always located directly at a repeater station $S$. A trial to establish an entangled pair for a repeater link is done by first creating an entangled pair, which takes a preparation time $T_\mathrm{P}$ (in practical terms this is often related to the the clock rate $f_\mathrm{clock}$ of the entangled pair source). After one of the qubits is loaded into the memory at station $S$, the other qubit is sent to a neighboring station $S^\prime$. Before any part of the entangled pair can be used, $S$ needs to wait for a classical message from station $S^\prime$ confirming whether the qubit arrived successfully. Therefore, one trial will take $t_\mathrm{trial}=T_\mathrm{P} + 2d/c$, where $d$ is the distance between $S$ and $S^\prime$, and $c = 2 \times 10^{8} \text{ m/s}$ is the speed of light in optical fibre.

This also means that when the pair is confirmed to be successfully established, the qubit at $S$ will already have been effected by memory noise for a duration of $2 d / c$ and the qubit at $S^\prime$ for $d/c$.

Since loss is a major source of imperfection, it is highly likely that multiple trials are needed to establish an entangled state between neighbouring stations.
If the trial is not successful the qubit in memory needs to be discarded and the process started again from the beginning.
The time it takes until the next entangled pair is established can be obtained by 
\begin{equation}
    k \times t_\mathrm{trial}
\end{equation}
where $k$ is drawn randomly from a geometric distribution $k(\eta_\mathrm{eff})$ with a success probability $\eta_\mathrm{eff}$. Note that both $t_\mathrm{trial}$ and $\eta_\mathrm{eff}$ are distance dependent.

In scenarios with multiple memories we assume the trials are performed simultaneously and independently for each memory slot. For simplicity we treat them as using spatially separate channels, although in practice one would likely need to consider allocation of certain time slots in the quantum channel to specific processes. However, as long as the time of one trial is much larger than the time it takes to prepare a new entangled pair for sending, any effects arising from this are very small.

\section{Results and Discussion}
At the core of an entanglement-swapping-based quantum repeater (as opposed to a scheme instead using quantum error correction \cite{Challenge1}) protocol lies the realization that subdividing a channel into multiple parts and using entanglement swapping can be beneficial. However, many previous investigations fall roughly into two very different categories.
On one side there are very abstract models that presume that a large ensemble of entangled states is readily available at the lowest repeater level and the main source of imperfections is the (potentially distance dependent) fidelity of these initial states, while the quantum memories are very high quality and the use of entanglement purification protocols is plentiful 
(see, \eg, Refs.~\cite{briegel1998quantum, Duer1999repeater, Zwerger2016, Wallnoefer2019}). In these kinds of scenarios choosing the number of 
purification steps at each repeater level is the main challenge in optimizing the protocols.

On the other side there are models that are closer to the experimental setups (see, \eg, 
Refs.~\cite{Collins2007, Luong2016, Shchukin2019, Trenyi2020}), where loss of qubits during the initial distribution of 
entangled states is a major factor and the coherence times of memories are rather short, and entanglement 
purification is hardly considered, if at all. Instead, taking into account the exact timing of qubits 
arriving and operations being performed is the focus of the analysis.
Remarkably, the basic approach of shorter channel segments and entanglement swapping has proved useful in these very different situations.

\subsection{Two links with entanglement purification}
\label{sec:twolink_epp}

\begin{figure*}
    \subfloat[]{
    \includegraphics[width=0.48\linewidth]{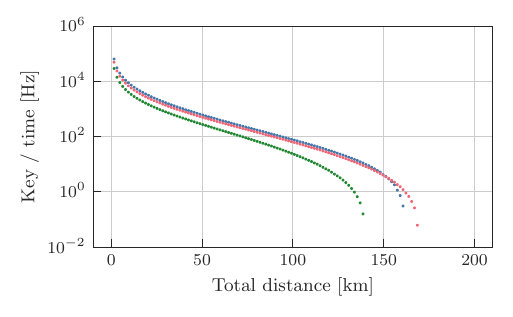}
    } \hfill \subfloat[]{
    \includegraphics[width=0.48\linewidth]{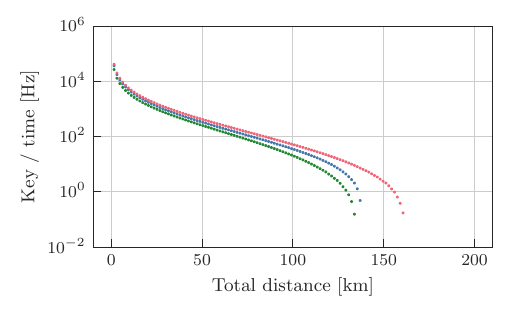}
    }
    \caption{\label{fig:twolink_epp}Obtainable key rates for two repeater links with 
    \textcolor{blue}{$0$ (blue)}, 
    \textcolor{red}{$1$ (red)} or 
    \textcolor{green}{$2$ (green)} 
    entanglement purification steps before entanglement swapping. (a) $F_\mathrm{init}=0.95$: Even though no \emph{entanglement purification protocol} (EPP) is necessary for most distances, 
    EPP can help extend the reachable distance. (b) $F_\mathrm{init}=0.935$: At worse initial fidelity 
    EPP can improve key rate for all distances.
    In both cases, it is also apparent that the number of purification steps needs 
    to be managed carefully. For these parameters, using two purification steps is actually detrimental.
    Other parameters: $T_\mathrm{dp} = 100 \text{ ms}$, $P_\mathrm{link} = 0.5$, 
    $p_\mathrm{d} = 10^{-6}$, $4$ memories per repeater link.}
\end{figure*}
In the following, we investigate a situation that can be seen as being located somewhere in between the two categories mentioned above: Quantum repeater protocols using entanglement purification while properly considering the timing considerations in the presence of realistic error models. 

It is easy to see the pros and cons of adding entanglement purification when considering the extreme ends of the parameter regime: 
If the initially generated entangled states are of such low fidelity that after performing entanglement swapping the output state is no longer useful (\eg,~it would introduce a too high quantum bit error rate for key distribution or fall short of some threshold fidelity at the output), no meaningful connection can be established without a mechanism to improve the fidelity. Entanglement purification protocols can help in this case by increasing the fidelity above the relevant threshold, if the quality of the quantum memories is sufficiently high to accommodate the additional time needed to perform them. However, if the decoherence time of the quantum memories is very short, 
clearly any gain in the fidelity from entanglement purification will long be gone by the time 
the required classical information arrives and therefore, performing entanglement swapping as soon as possible is more desirable.

Here we investigate parameters in between those extremes and especially initial fidelities $F_\mathrm{init}$ for which both approaches with and without the use of an \textit{entanglement purification protocol} (EPP) can, in principle, achieve non-zero key rates. This allows us to quantify under which circumstances the use of EPPs can be beneficial in this intermediate regime and what memory times need to be achieved in order to make that possible. A similar perspective on the trade-off between using additional memories for entanglement purification or multiplexing has been studied in Ref.~\cite{Bernardes2011}, albeit with perfect quantum memories.

For now, consider a setup with a single repeater station, \ie, consisting of two repeater links. The repeater station is equipped with an entangled pair source and four memories per repeater link, so multiple pairs can potentially be established simultaneously. Either 0, 1 or 2 steps of the DEJMPS entanglement purification protocol
 \cite{dejmps} (see also Appendix \ref{sec:explain_epp}) are performed on each side before the station applies entanglement swapping. The established connections are then used to perform quantum key distribution. However, entanglement purification is a probabilistic process and  measurement outcomes have to be communicated between the involved parties in order for them to know whether the EPP has been successful.

If no entanglement purification is performed, the end stations do not need to have quantum memories, because the arriving qubits can be measured right away. However, when performing entanglement purification the end stations will also need to be equipped with quantum memories. 
Since in this instance we are only concerned about the secret key rate and not about storing a final long distance entangled state, at the end stations the output qubits of the final EPP step can be measured immediately to avoid additional time in memory, even though the classical information about whether the entanglement purification was successful has not arrived yet.

In Fig.~\ref{fig:twolink_epp},
the achievable key rates with varying number of entanglement purification steps 
are shown for a selection of specific values of $F_\mathrm{init}$, for which the use of entanglement purification is just barely advantageous. For the given parameters, at  $F_\mathrm{init} = 0.95$ using entanglement purification actually leads to a slight reduction in the achievable key rate for most distances (e.g. key rates of \textasciitilde61 Hz with one step of EPP compared to \textasciitilde75 Hz without EPP at 100 km). However, the use of EPP can extend the distances, for which a positive key rate can be achieved, by a bit (from \textasciitilde160 km to \textasciitilde170 km), because there the reduction in error rate by the EPP is actually worth the lower raw bit rate. As expected from the extreme parameter regimes discussed above, if $F_\mathrm{init}$ is sufficiently low, the use of entanglement purification indeed is just outright beneficial (as is the case in Fig.~\ref{fig:twolink_epp}b with $F_\mathrm{init} = 0.935$). 

For both of these cases, it is also clear that doing more than one step of entanglement purification is not helpful. However, more purification steps may become useful at even lower $F_\mathrm{init}$. In any case, clearly the number of purification steps to be performed as part of the protocol needs to be carefully adjusted according to which experimental parameters are available.

We remark that in order to optimize this even further other schemes of entanglement purification could be considered. For example, a pumping scheme \cite{Duer2003} instead of a strict repetition scheme could be employed. Alternatively, the number of entanglement purification steps could be dynamically adjusted depending on whether the other link is already ready for entanglement swapping or not.

\begin{figure}
    \centering
    \includegraphics[width=\linewidth]{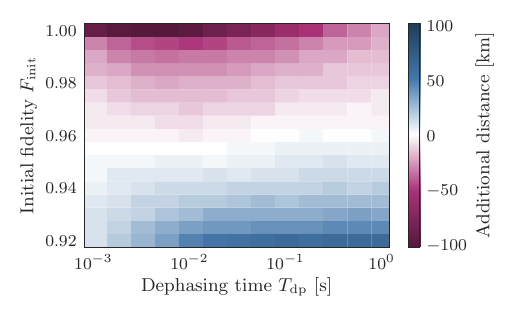}
    \caption{\label{fig:twolink_epp_2d_plot} Increase in achievable distance (\ie~with non-zero key rate) using a repeater protocol with one entanglement purification step compared to a repeater protocol without entanglement purification.
    Other parameters: $P_\mathrm{link}=0.5$, $p_d=10^{-6}$, $4$ memories per repeater link.
    }
\end{figure}

While the above examples demonstrate there are situations for which the use of EPPs is worth considering, in Fig.~\ref{fig:twolink_epp_2d_plot} a wider range of parameters is considered. It shows the increase in reachable distance, \ie~where a non-zero key rate can be still be achieved, when using one EPP step compared to not using entanglement purification. As expected, the addition of entanglement purification is most impactful when $F_\mathrm{init}$ is low and the memory quality (represented by dephasing time $T_\mathrm{dp}$) is high. However, it also shows that the usefulness is not restricted to the most extreme cases, but the addition of entanglement purification can somewhat extend secure transmission distances for a wide range of parameters.

\subsubsection{Cut-off time as an alternative strategy}
In essence, the use of entanglement purification can be understood as one way of trading some of the raw rate for a higher fidelity of the pairs that are used. However, another strategy that is also reducing the raw rate in exchange for a lower error rate, is simply discarding qubits that have been kept in storage for too long. A common mechanism for this is to choose a fixed cutoff time $t_\mathrm{cut}$ (see, \eg, Ref.~\cite{Rozpedek2018}) after which a qubit is discarded, and the process of establishing a 
link is started again. In general, optimizing whether to keep or discard an existing link has been shown to be exponentially hard \cite{Khatri2021} since all decisions at previous times need to be taken into account as well to formulate the optimal strategy. However, at the individual link level a cut-off time strategy is indeed optimal in the steady state limit \cite{Khatri2022}. 

\begin{figure}
    \centering
    \includegraphics[width=\linewidth]{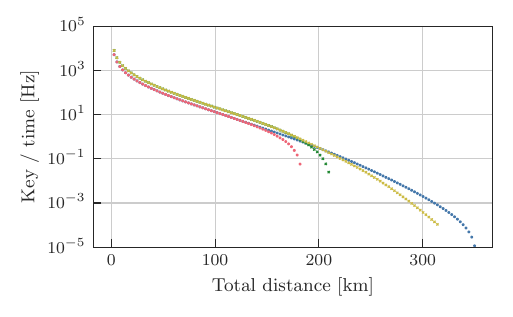}
    \caption{Achievable key rates for two repeater links with a combination of strategies. Without entanglement purification and cut-off times 
    \textcolor{red}{$t_\mathrm{cut} = \infty$ (red)} or 
    \textcolor{blue}{$50 \text{ ms}$ (blue)}. 
    Alternatively, a protocol with one step of entanglement purification and cut-off times 
    \textcolor{green}{$t_\mathrm{cut} = \infty$ (green)} or 
    \textcolor{yellow}{$650 \text{ ms}$ (yellow)}.
    Properly optimized cut-off times can clearly help in both cases.
    Other parameters: $T_\mathrm{dp}=1 \text{ s}$, $P_\mathrm{link}=0.5$, $F_\mathrm{init} = 0.925$, 2 memories per repeater link.}
    \label{fig:twolink_epp_compare_cutoff}
\end{figure}

We applied the cut-off time approach both for the protocol with and without EPP for an $F_\mathrm{init}$ that is just barely above the threshold where a non-EPP protocol is viable. A selection of the best cut-off times we found for both approaches is shown in 
Fig.~\ref{fig:twolink_epp_compare_cutoff}. First, it shows that also the EPP approach can benefit from some upper limit on the storage time. Naturally, the optimal cut-off time is much higher for the entanglement purification approach, as one must allow time for the classical information after the entanglement purification as well. This demonstrates that one necessarily needs to consider combinations of approaches when designing quantum repeater protocols.
Furthermore, this shows the trade-off between the EPP and cut-off time strategies is complex: for short distances the use of entanglement purification leads to higher key rates, however, the simple protocol without entanglement purification (but with optimized cut-off times) is the better choice at long distances for this parameter set.

Even though entanglement purification and cut-off times are used for similar reasons, it should be noted there is a fundamental difference between them. While cut-off times can reduce the effect of decoherence on the final states, it cannot increase the fidelity of an individual state. For even lower $F_\mathrm{init}$, where using a method to increase the fidelity of a pair (like EPPs) is necessary, simply optimizing the cut-off time cannot be sufficient.

\subsection{Multiple repeater links} \label{sec:multilink}
Already a single repeater station is very useful in reaching longer distances and allows the fundamental limits of repeaterless quantum communication to be overcome. However, with the increasing loss along distances, invariably one will need to consider adding more repeater stations at some point, \eg,~the exponential loss in optical fibres is a severe limitation when aiming for intercontinental distances.
Indeed, the multiple station scenario is where the quantum memories truly become indispensable, as for a single station so-called twin-field QKD \cite{Lucamarini2018} achieves repeater-like scaling with only a simplified relay station that can perform measurements but has no memories.

In principle, subdividing a channel into more segments is very attractive. After all, if adding in a repeater station can improve the connection between two parties under certain conditions, it stands to reason that the same principle can be applied for each individual link. For each individual segment the channel loss in much shorter connections will be lower. Naturally, this does not come without a trade-off: All the imperfections that are not distance dependent but instead are related to generation of the entangled states, handling the states and performing operations, will be present at each of these repeater links, therefore affecting the output multiple times.
In order to optimize the connection between two distant parties, the number of repeater links is an additional factor to consider and will certainly depend on the parameters of the available quantum hardware. 

For repeater setups with multiple repeater links (without entanglement purification) some expressions for average waiting times are known, \eg,~with probabilistic entanglement swapping \cite{Shchukin2019, Collins2007, Inesta22} or for specific loss models, \eg,~suited for satellite-based repeaters \cite{Gundogan2021}. For some setups even expressions for the obtainable key have been found recently \cite{Kamin2022}.

\begin{figure*}
    \subfloat[]{
    \includegraphics[width=\columnwidth]{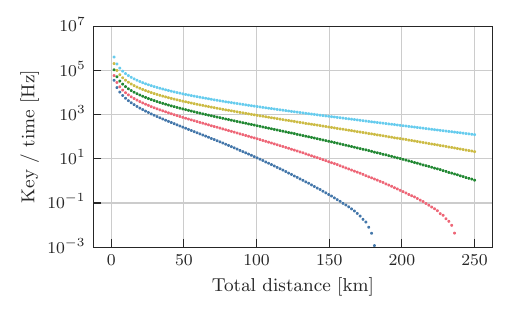}
    } \hfill \subfloat[]{
    \includegraphics[width=\columnwidth]{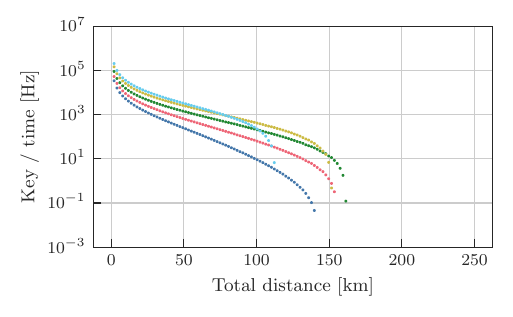}
    }
    \caption{\label{fig:manylink} Obtainable key rates
    when using protocols with 
    \textcolor{blue}{$2$ (blue)}, 
    \textcolor{red}{$4$ (red)}, 
    \textcolor{green}{$8$ (green)}, 
    \textcolor{yellow}{$16$ (yellow)} or 
    \textcolor{cyan}{$32$ (cyan)} 
    repeater links. 
    (a) With $F_\mathrm{init} = 1.0$ utilizing more repeater links is generally favorable. (b) Even with small constant overheads per repeater link ($F_\mathrm{init} = 0.998$) trade-offs become apparent.
    Other parameters:
    $T_\mathrm{dp} = 10 \text{ ms}$,
    $P_\mathrm{link} = 0.5$,
    $p_\mathrm{d} = 10^{-6}$.
    }
\end{figure*}

To investigate the inherent trade-off when adding additional repeater stations mentioned above we opt for a simple model of fixed, per-link overhead: We consider an imperfect initial fidelity $F_\mathrm{init} < 1.0$ of the generated entangled states at each repeater link. This could be interpreted as either imperfect entangled pair sources or some additional constant imperfections that arise from handling the states at the repeater station, \eg,~by the read-in procedure that loads an arriving qubit into the quantum memory. One could also consider different sources of imperfections that happen at every link, such as imperfect operations to perform Bell state measurements at the repeater stations (which has a similar effect as lowering $F_\mathrm{init}$), or even a combination of multiple noisy processes. However, we focus on a single parameter for now, in order to keep the results easier to interpret. Furthermore, we limit our investigation to an equidistant spacing of repeater stations, although this may not be optimal in some cases, \eg,~when photons cannot be sent in both directions simultaneously \cite{Luong2016}, and asymmetric repeater setups come with their own set of considerations, see, \eg, Ref.~\cite{Wallnoefer2020}.
In Fig.~\ref{fig:manylink}a, perfect initial states 
($F_\mathrm{init}=1.0$) with multiple repeater stations are 
considered. While there are still some factors to consider that arise from using 
multiple repeater stations, such as many qubits potentially dephasing in quantum memories at multiple stations, adding more repeater links is generally favorable. E.g.~at \textasciitilde150 km key rates are improved by a factor of approximately 30, 260, 1100 or 3500 when using 4, 8, 16 or 32 repeater links, respectively, compared to a two-link approach.

However, adding even a tiny constant overhead per repeater link immediately demonstrates that there is a serious trade-off when adding more repeater stations, as is apparent from Fig.~\ref{fig:manylink}b with $F_\mathrm{init}=0.998$. It also clearly illustrates that the optimal number of repeater stations can be different depending on the distance between the end stations, \eg,~for these parameters using a protocol with $32$ repeater links leads to improved key rates at $50$~km distance, but the protocol with $8$ repeater links allows to reach much longer distances overall.

\begin{figure}
    \centering
    \includegraphics[width=\linewidth]{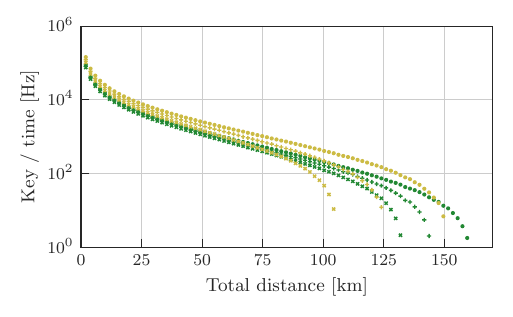}
    \caption{\label{fig:manylink_improve_f}
    Key rates for 
    \textcolor{green}{$8$ (green)} or 
    \textcolor{yellow}{$16$ (yellow)} 
    repeater links, with varying initial fidelities: $F_\mathrm{init} = 0.996 
    \text{ (crosses), } 0.997 
    \text{ (plusses), } 0.998 
    \text{ (circles)}$.
    Improving some experimental parameters can enable alternative strategies.
    E.g.,~when connecting end stations 125 km apart, improving $F_\mathrm{init}$ 
    from $0.996$ to $0.998$ allows changing from a 8-link protocol to a 16-link protocol, 
    which improves the obtainable key rate more than the increase in $F_\mathrm{init}$ alone.
    Other parameters:
    $T_\mathrm{dp} = 10 \text{ ms}$,
    $P_\mathrm{link} = 0.5$,
    $p_\mathrm{d} = 10^{-6}$.
    }
\end{figure}

Naturally, the point where adding more repeater stations becomes detrimental depends strongly on the per-link overhead itself. When looking at different values of $F_\mathrm{init}$ in Fig.~\ref{fig:manylink_improve_f}, 
we can see that making efforts to improve an experimental parameter can have implications beyond simply increasing achievable rates for a fixed strategy. For example,~when considering end stations 125~km apart for this particular set of parameters, improving the achievable $F_\mathrm{init}$ allows us to switch to using a protocol with more repeater links that increases the obtainable key rate even more than the $F_\mathrm{init}$ improvement alone would have done.

This demonstrates that considering multiple different repeater strategies is vital when analyzing quantum communication setups. Even if more complex schemes are not realistically possible at present, if improved operations or devices are likely to exist in the future, it might open possibilities for new strategies. Furthermore, this example illustrates how simulation tools can potentially guide experimental developments. Being able to quantify the effect of improving a certain experimental parameter is essential to make an informed decision on where to focus future efforts.

\subsubsection{Entanglement purification at the lowest level}\label{sec:manylink_epp}

When considering entanglement purification as an additional tool for repeater protocols with multiple links, many variations of protocols are possible (\eg~an approach that is generally similar but differs in detail is analyzed in Ref.~\cite{NetSquid}). This is because in principle one can apply entanglement purification protocols at multiple points of the whole process, \eg,~after a certain number of swapping operations it can even become essential to improve the fidelity of the entangled pair via entanglement purification lest the systems becomes completely disentangled. In fact, due to the relative increase in fidelity of repetition protocols being dependent on the input states, on which repeater levels to put the entanglement purification steps has been a consideration for entanglement-swapping-based quantum repeaters from the very beginning \cite{briegel1998quantum, Duer1999repeater}. However, with coherence times of quantum memories remaining a central limitation for the foreseeable future, it remains doubtful how useful entanglement purification at higher repeater levels can be. One entanglement purification step at that level is costly in terms of resources as for the creation of each of the long distance pairs multiple lowest level pairs have been used. Furthermore, with 
the increasing distance at higher repeater levels, the transmission of the classical information necessary for performing the entanglement purification protocol will take longer and at some point the increase 
in fidelity achieved through a 
entanglement purification protocol will be negated by the noise processes in the memory during this time.

Optimizing repeater protocols with entanglement purification in mind therefore inevitably means considering these complex trade-offs. For the purpose of this work we only consider whether adding entanglement purification at the lowest level, \ie,~purifying only the initial entangled pairs distributed to neighboring repeater stations, can already be beneficial. 

\begin{figure}
    \centering
    \includegraphics[width=\linewidth]{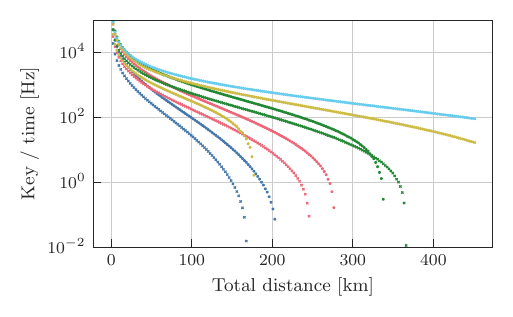}
    \caption{\label{fig:manylink_epp} Entanglement purification can extend achievable ranges for quantum repeaters with multiple links. Key rates without entanglement purification (circles) and with one step of entanglement purification at the lowest level (crosses) when using 
    \textcolor{blue}{$2$ (blue)}, 
    \textcolor{red}{$4$ (red)}, 
    \textcolor{green}{$8$ (green)}, 
    \textcolor{yellow}{$16$ (yellow)} or 
    \textcolor{cyan}{$32$ (cyan)} 
    repeater links. 
    Parameters:
    $T_\mathrm{dp} = 100 \text{ ms}$,
    $F_\mathrm{init} = 0.99$,
    $P_\mathrm{link} = 0.5$,
    $p_\mathrm{d} = 10^{-6}$, $2$ quantum memories per repeater link at each station.
    }
\end{figure}

We consider a setup with high quality memories ($T_\mathrm{dp} = 100 \text{ ms}$), which favors entanglement purification, and analyse the question of whether one step of entanglement purification at the lowest level should be performed. The results are depicted in Fig.~\ref{fig:manylink_epp}; they show that for this parameter set entanglement purification significantly changes the impact of the trade-offs being made when including additional repeater stations. In fact, it is what makes using more repeater stations viable in this case, which in turn improves the reachable distances significantly.
This is one example how the use of entanglement purification can be beneficial not just when considering very low initial fidelities, but even when dealing with initial states with relatively high $F_\mathrm{init}$.
More generally, it showcases how how the access to certain tools (like entanglement purification) makes it necessary to reevaluate other aspects of how the repeater setup is constructed.

\subsection{Impact of improving parameters for custom error models}
\label{sec:many_params}
For the previous results, we have used a fairly uniform error model to make the results easy to interpret. In order to showcase a broader range of error models, we consider an alternative noise model and analyze the effect of improving certain parameters of the hardware. This also follows the line of thought of the previous scenario that a small change in conditions can have a significant effect on the overall outcome.

We consider quantum memories with a time-dependent amplitude damping channel captured as
\begin{equation}
    \begin{aligned}
    \mathcal{E}_\mathrm{damp}^{(i)}(t) \rho = 
    \begin{pmatrix}
    1 & 0 \\
    0 & \sqrt{1 - \gamma(t)}
    \end{pmatrix} &\rho \begin{pmatrix}
    1 & 0 \\
    0 & \sqrt{1 - \gamma(t)}
    \end{pmatrix} \\
    + \begin{pmatrix}
        0 & \sqrt{\gamma(t)} \\
        0 & 0
    \end{pmatrix} &\rho \begin{pmatrix}
        0 & 0 \\
        \sqrt{\gamma(t)} & 0
    \end{pmatrix}
    \end{aligned}
\end{equation}
with $\gamma(t) = 1 - e^{-t / T_\mathrm{damp}}$ and the damping time $T_\mathrm{damp}$.
Furthermore, we assume that both entanglement purification and
Bell state measurements are performed in a gate based way with the
same two-qubit gate error parameter $p_\mathrm{gate}$. The noisy CNOT gates involved in both of these operations are modeled as local depolarizing noise channels $\mathcal{E}_\mathrm{d}(p_\mathrm{gate})$ acting on both of the input qubits, followed by the perfect gate operation.
The local depolarizing channel on the $i$-th qubit is defined as
\begin{equation}
    \mathcal{E}_\mathrm{d}^{(i)}(p_\mathrm{gate}) \rho = 
    p_\mathrm{gate} \rho + \frac{1-p_\mathrm{gate}}{2} \mathrm{tr}_i(\rho) \otimes \mathbbm{1}^{(i)}
\end{equation}
with $\mathrm{tr}_i$ denoting the partial trace over the $i$-th subsystem.

\begin{figure*}
    \subfloat[]{
    \includegraphics[width=0.48\linewidth]{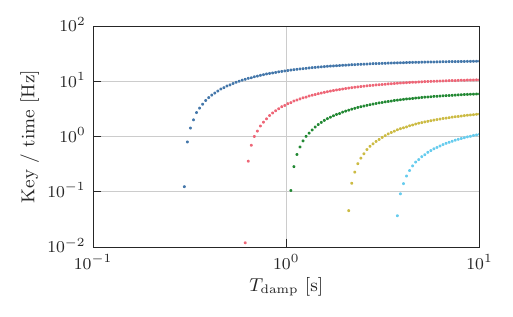}
    } \hfill \subfloat[]{
    \includegraphics[width=0.48\linewidth]{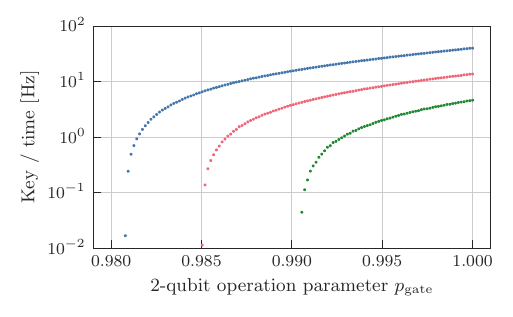}
    }
    \caption{\label{fig:many_params} The impact of improving certain hardware parameters for a custom error model. Using a repeater protocol with 4 repeater links and 1 entanglement purification step at the lowest repeater level. 2 quantum memories per connected link are available at each station. Parameters: $P_\mathrm{link}=0.01$,
        $p_\mathrm{d}=10^{-6}$,
        $F_\mathrm{init}= 0.99$.
    For a total total distance of
    \textcolor{blue}{$20$ (blue)}, 
    \textcolor{red}{$35$ (red)}, 
    \textcolor{green}{$50$ (green)}, 
    \textcolor{yellow}{$75$ (yellow)} or 
    \textcolor{cyan}{$100$ (cyan)} km
    (a) varying memory quality with fixed $p_\mathrm{gate}=0.99$ ;
    (b) varying quality of 2-qubit gates with fixed $T_\mathrm{damp}=1\text{s}$ (no positive key rate for $75$ or $100$ km).
    }
\end{figure*}

In Fig.~\ref{fig:many_params}, the impact of making improvements to either the memories (higher $T_\mathrm{damp}$) or 2-qubit gates (higher $p_\mathrm{gate}$) on the reachable key rate is shown for a protocol with $4$ repeater links and one entanglement purification step at the lowest level. It is clear that both parameters have strict minimum requirements to achieve a non-zero key rate with this protocol at a specified distance.

\section{Related work}
\label{sec:related_work}
In recent years many tools have been released to study various aspects of quantum repeaters and quantum networks. Naturally, each of them focus on different aspects of quantum communication and use differing assumptions and abstractions to model what is important for their specific considerations. Such a variety of interpretations and implementations is certainly beneficial to the discussion in the field and the overall understanding of quantum networks. At the same time, the field of quantum network simulation is not really mature enough to have a standardized framework for comparing features and protocols. In the following we give a brief overview over some of the available simulation platforms.

\emph{SimulaQron} \cite{simulaqron} provides a distributed classical simulation of multiple connected parties with quantum computers that allows to use the delay of real world classical networks as part of their model. Furthermore, it allows the development of software for quantum networks based on the instruction set architecture \emph{NetQASM}  \cite{NetQASM}, which would then allow the protocols that were built using the simulation to be run on quantum hardware.

\emph{QuISP} \cite{satoh2021quisp} uses an event-driven approach to simulate quantum networks, but it places a much stronger focus on the networking aspects for large-scale networks. It achieves the necessary scalability by updating an error model rather than tracking full quantum states.

Close in mindset to the approach introduced in this work are
the approaches followed by \emph{NetSquid}  \cite{NetSquid} and \emph{SeQUeNCe} \cite{sequence}, which are both event-based software tools for modelling and simulating quantum networks.
\emph{NetSquid} has also been used in the context of quantum repeater chains with realistic hardware. In Ref.~\cite{NetSquid}, a setup with a different variation of an entanglement purification strategy is analyzed for a NV-center based quantum repeater stations with additional hardware restrictions. It has also used to consider aspects for quantum repeaters based on atomic ensembles \cite{thesis_maier} and real world fiber infrastructure \cite{Avis22}, as well as the use of genetic algorithms to optimize repeater protocols \cite{thesis_Labay_Mora, Ferreira_da_Silva_2021}.
\emph{SeQUeNCe} has investigated the distribution of entangled states in a nine node network in Ref.~\cite{sequence}. It has also been used to model and analyze a specific generation procedure for photonic quantum states with absorptive quantum memories \cite{sequence_absorptive}.
In Ref.~\cite{sequence_parallel}, it is explained how the simulator can parallelise some calculations.

\section{Summary and outlook}

Quantum communication constitutes a central pillar of quantum technologies, with 
cryptographic applications being the main practical application. The vision of achieving secure communication 
over arbitrary distances requires overcoming the strict limitations that stem from not having quantum repeaters available.
Hence, the quest for achieving realistic and efficient quantum repeater protocols is by no means a detail, but is actually a core step before
practically relevant protocols of quantum communication over arbitrary distances become viable. 
Research on hardware development is progressing well, but is challenged by the situation that the
multitude of design rules render the optimal construction of quantum repeaters challenging. 

In order to substantially assist this task, we have developed a numerical simulation that allows us to analyse 
practical quantum repeater setups, as a versatile scheme, but detailed enough that a multitude of
physical aspects can be accommodated. We stress that our approach can obtain not only waiting times but also calculate quantities directly from the output distribution (such as secret key rates) while taking into account the probabilistic distribution of initial states, time-dependent memory noise, multiple quantum memories per repeater link, entanglement purification and multiple repeater links (including analyses for up to 32 in this work and a demonstration of up to 1024 in Appendix \ref{sec:appendix_scaling}).

 We have analyzed multiple scenarios that are focused on the comparison and combination of repeater strategies:
 We have explored the use of entanglement purification as part of a quantum repeater protocol and quantified under which circumstances employing such a strategy can be beneficial. More importantly, we found that increasing the complexity in this way may be unwarranted if a strategy with optimized cut-off times can produce similar or even greater performance.
 Furthermore, we have analyzed the trade-offs that come with protocols that use multiple repeater links and provided a clear example how improving the experimental capabilities (\eg~better initial fidelities) can make new options for optimisation (\eg~ using more repeater stations) viable, which leads to a larger improvement together than what would be obtainable in isolation.

 In general, there are a multitude of questions when it comes to the design of quantum repeaters or, at a  higher level, quantum networks. One of the main challenges in exploring further options is that the parameter space is not only vast, but any result of optimality can only be understood in a specific context and with a specific set of tools in mind. While our approach will certainly allow to explore different parameter regimes in a more rigorous manner, 
 we expect it will be particularly useful for investigating advanced repeater protocols, e.g., 
 for asymmetric setups or with dynamic adjustments of strategies. 
 Another possible direction is to use a numerical simulation to closely model particular experimental 
 setups and investigate the viability of building more complex schemes with currently available experimental hardware (\eg, Ref.~\cite{Avis22} follows such an approach). It is the hope that the versatile platform presented here substantially
 contributes to the quest of identifying good design principles for feasible quantum
 repeaters and hence contributions to achieving quantum communication over arbitrary
 distances.\\ 

\section{Acknowledgments}
J.~W.~and J.~E.~acknowledge support from the Federal 
Ministry of Education and 
Research (BMBF) 
via projects Q.Link.X and QR.X. 
F.~W.~and F. H.~acknowledge support from the Emmy Noether DFG 
grant No.~418294583.
N.~W. and J.~E.~acknowledge support from the Federal Ministry of Education and Research (BMBF) via the projects PhoQuant and QPIC-1.
J.~E.~also thanks the European Research Council (DebuQC) for support.
We thank the HPC Service of ZEDAT \cite{hpc_fub}, Freie Universität Berlin, and the HPC Service of the Department of Physics, Freie Universität Berlin, for computing time.
 
\section{Code availability}
The simulation framework \emph{\mbox{ReQuSim}} \cite{requsim} is available as an open source Python package.
The code that has been used to generate all results presented this work is archived at \mbox{DOI: \href{https://doi.org/10.5281/zenodo.7399897}{10.5281/zenodo.7399897}}.
It uses a combination of \emph{\mbox{ReQuSim} v0.4} and an earlier version of the code base that has not been released as a stand-alone package.
The raw output data of the simulation that has been used to calculate the values presented in the plots is available upon reasonable request.
 
\bibliographystyle{apsrev4-2}
\bibliography{repeater_simulation.bib}

\begin{thebibliography}{60}%
\makeatletter
\providecommand \@ifxundefined [1]{%
 \@ifx{#1\undefined}
}%
\providecommand \@ifnum [1]{%
 \ifnum #1\expandafter \@firstoftwo
 \else \expandafter \@secondoftwo
 \fi
}%
\providecommand \@ifx [1]{%
 \ifx #1\expandafter \@firstoftwo
 \else \expandafter \@secondoftwo
 \fi
}%
\providecommand \natexlab [1]{#1}%
\providecommand \enquote  [1]{``#1''}%
\providecommand \bibnamefont  [1]{#1}%
\providecommand \bibfnamefont [1]{#1}%
\providecommand \citenamefont [1]{#1}%
\providecommand \href@noop [0]{\@secondoftwo}%
\providecommand \href [0]{\begingroup \@sanitize@url \@href}%
\providecommand \@href[1]{\@@startlink{#1}\@@href}%
\providecommand \@@href[1]{\endgroup#1\@@endlink}%
\providecommand \@sanitize@url [0]{\catcode `\\12\catcode `\$12\catcode
  `\&12\catcode `\#12\catcode `\^12\catcode `\_12\catcode `\%12\relax}%
\providecommand \@@startlink[1]{}%
\providecommand \@@endlink[0]{}%
\providecommand \url  [0]{\begingroup\@sanitize@url \@url }%
\providecommand \@url [1]{\endgroup\@href {#1}{\urlprefix }}%
\providecommand \urlprefix  [0]{URL }%
\providecommand \Eprint [0]{\href }%
\providecommand \doibase [0]{https://doi.org/}%
\providecommand \selectlanguage [0]{\@gobble}%
\providecommand \bibinfo  [0]{\@secondoftwo}%
\providecommand \bibfield  [0]{\@secondoftwo}%
\providecommand \translation [1]{[#1]}%
\providecommand \BibitemOpen [0]{}%
\providecommand \bibitemStop [0]{}%
\providecommand \bibitemNoStop [0]{.\EOS\space}%
\providecommand \EOS [0]{\spacefactor3000\relax}%
\providecommand \BibitemShut  [1]{\csname bibitem#1\endcsname}%
\let\auto@bib@innerbib\@empty
%</preamble>
\bibitem [{\citenamefont {Acin}\ \emph {et~al.}(2018)\citenamefont {Acin},
  \citenamefont {Bloch}, \citenamefont {Buhrman}, \citenamefont {Calarco},
  \citenamefont {Eichler}, \citenamefont {Eisert}, \citenamefont {Esteve},
  \citenamefont {Gisin}, \citenamefont {Glaser}, \citenamefont {Jelezko},
  \citenamefont {Kuhr}, \citenamefont {Lewenstein}, \citenamefont {Riedel},
  \citenamefont {Schmidt}, \citenamefont {Thew}, \citenamefont {Wallraff},
  \citenamefont {Walmsley},\ and\ \citenamefont {Wilhelm}}]{Roadmap}%
  \BibitemOpen
  \bibfield  {author} {\bibinfo {author} {\bibfnamefont {A.}~\bibnamefont
  {Acin}}, \bibinfo {author} {\bibfnamefont {I.}~\bibnamefont {Bloch}},
  \bibinfo {author} {\bibfnamefont {H.}~\bibnamefont {Buhrman}}, \bibinfo
  {author} {\bibfnamefont {T.}~\bibnamefont {Calarco}}, \bibinfo {author}
  {\bibfnamefont {C.}~\bibnamefont {Eichler}}, \bibinfo {author} {\bibfnamefont
  {J.}~\bibnamefont {Eisert}}, \bibinfo {author} {\bibfnamefont
  {J.}~\bibnamefont {Esteve}}, \bibinfo {author} {\bibfnamefont
  {N.}~\bibnamefont {Gisin}}, \bibinfo {author} {\bibfnamefont {S.~J.}\
  \bibnamefont {Glaser}}, \bibinfo {author} {\bibfnamefont {F.}~\bibnamefont
  {Jelezko}}, \bibinfo {author} {\bibfnamefont {S.}~\bibnamefont {Kuhr}},
  \bibinfo {author} {\bibfnamefont {M.}~\bibnamefont {Lewenstein}}, \bibinfo
  {author} {\bibfnamefont {M.~F.}\ \bibnamefont {Riedel}}, \bibinfo {author}
  {\bibfnamefont {P.~O.}\ \bibnamefont {Schmidt}}, \bibinfo {author}
  {\bibfnamefont {R.}~\bibnamefont {Thew}}, \bibinfo {author} {\bibfnamefont
  {A.}~\bibnamefont {Wallraff}}, \bibinfo {author} {\bibfnamefont
  {I.}~\bibnamefont {Walmsley}},\ and\ \bibinfo {author} {\bibfnamefont
  {F.~K.}\ \bibnamefont {Wilhelm}},\ }\href
  {https://doi.org/10.1088/1367-2630/aad1ea} {\bibfield  {journal} {\bibinfo
  {journal} {New J. Phys.}\ }\textbf {\bibinfo {volume} {20}},\ \bibinfo
  {pages} {080201} (\bibinfo {year} {2018})}\BibitemShut {NoStop}%
\bibitem [{\citenamefont {Gisin}\ \emph {et~al.}(2002)\citenamefont {Gisin},
  \citenamefont {Ribordy}, \citenamefont {Tittel},\ and\ \citenamefont
  {Zbinden}}]{RevModPhys.74.145}%
  \BibitemOpen
  \bibfield  {author} {\bibinfo {author} {\bibfnamefont {N.}~\bibnamefont
  {Gisin}}, \bibinfo {author} {\bibfnamefont {G.}~\bibnamefont {Ribordy}},
  \bibinfo {author} {\bibfnamefont {W.}~\bibnamefont {Tittel}},\ and\ \bibinfo
  {author} {\bibfnamefont {H.}~\bibnamefont {Zbinden}},\ }\href
  {https://doi.org/10.1103/RevModPhys.74.145} {\bibfield  {journal} {\bibinfo
  {journal} {Rev. Mod. Phys.}\ }\textbf {\bibinfo {volume} {74}},\ \bibinfo
  {pages} {145} (\bibinfo {year} {2002})}\BibitemShut {NoStop}%
\bibitem [{\citenamefont {Pirandola}\ \emph {et~al.}(2020)\citenamefont
  {Pirandola}, \citenamefont {Andersen}, \citenamefont {Banchi}, \citenamefont
  {Berta}, \citenamefont {Bunandar}, \citenamefont {Colbeck}, \citenamefont
  {Englund}, \citenamefont {Gehring}, \citenamefont {Lupo}, \citenamefont
  {Ottaviani}, \citenamefont {Pereira}, \citenamefont {Razavi}, \citenamefont
  {Shaari}, \citenamefont {Tomamichel}, \citenamefont {Usenko}, \citenamefont
  {Vallone}, \citenamefont {Villoresi},\ and\ \citenamefont
  {Wallden}}]{NewQuantumCryptoReview}%
  \BibitemOpen
  \bibfield  {author} {\bibinfo {author} {\bibfnamefont {S.}~\bibnamefont
  {Pirandola}}, \bibinfo {author} {\bibfnamefont {U.~L.}\ \bibnamefont
  {Andersen}}, \bibinfo {author} {\bibfnamefont {L.}~\bibnamefont {Banchi}},
  \bibinfo {author} {\bibfnamefont {M.}~\bibnamefont {Berta}}, \bibinfo
  {author} {\bibfnamefont {D.}~\bibnamefont {Bunandar}}, \bibinfo {author}
  {\bibfnamefont {R.}~\bibnamefont {Colbeck}}, \bibinfo {author} {\bibfnamefont
  {D.}~\bibnamefont {Englund}}, \bibinfo {author} {\bibfnamefont
  {T.}~\bibnamefont {Gehring}}, \bibinfo {author} {\bibfnamefont
  {C.}~\bibnamefont {Lupo}}, \bibinfo {author} {\bibfnamefont {C.}~\bibnamefont
  {Ottaviani}}, \bibinfo {author} {\bibfnamefont {J.}~\bibnamefont {Pereira}},
  \bibinfo {author} {\bibfnamefont {M.}~\bibnamefont {Razavi}}, \bibinfo
  {author} {\bibfnamefont {J.~S.}\ \bibnamefont {Shaari}}, \bibinfo {author}
  {\bibfnamefont {M.}~\bibnamefont {Tomamichel}}, \bibinfo {author}
  {\bibfnamefont {V.~C.}\ \bibnamefont {Usenko}}, \bibinfo {author}
  {\bibfnamefont {G.}~\bibnamefont {Vallone}}, \bibinfo {author} {\bibfnamefont
  {P.}~\bibnamefont {Villoresi}},\ and\ \bibinfo {author} {\bibfnamefont
  {P.}~\bibnamefont {Wallden}},\ }\href {https://doi.org/10.1364/AOP.361502}
  {\bibfield  {journal} {\bibinfo  {journal} {Adv. Opt. Photon.}\ }\textbf
  {\bibinfo {volume} {12}},\ \bibinfo {pages} {1012} (\bibinfo {year}
  {2020})}\BibitemShut {NoStop}%
\bibitem [{\citenamefont {Hillery}\ \emph {et~al.}(1999)\citenamefont
  {Hillery}, \citenamefont {Bu{\v z}ek},\ and\ \citenamefont
  {Berthiaume}}]{Hillery:1999tb}%
  \BibitemOpen
  \bibfield  {author} {\bibinfo {author} {\bibfnamefont {M.}~\bibnamefont
  {Hillery}}, \bibinfo {author} {\bibfnamefont {V.}~\bibnamefont {Bu{\v
  z}ek}},\ and\ \bibinfo {author} {\bibfnamefont {A.}~\bibnamefont
  {Berthiaume}},\ }\href {https://doi.org/10.1103/PhysRevA.59.1829} {\bibfield
  {journal} {\bibinfo  {journal} {Phys. Rev. A}\ }\textbf {\bibinfo {volume}
  {59}},\ \bibinfo {pages} {1829} (\bibinfo {year} {1999})}\BibitemShut
  {NoStop}%
\bibitem [{\citenamefont {Murta}\ \emph {et~al.}(2020)\citenamefont {Murta},
  \citenamefont {Grasselli}, \citenamefont {Kampermann},\ and\ \citenamefont
  {Bruss}}]{Murta:2020de}%
  \BibitemOpen
  \bibfield  {author} {\bibinfo {author} {\bibfnamefont {G.}~\bibnamefont
  {Murta}}, \bibinfo {author} {\bibfnamefont {F.}~\bibnamefont {Grasselli}},
  \bibinfo {author} {\bibfnamefont {H.}~\bibnamefont {Kampermann}},\ and\
  \bibinfo {author} {\bibfnamefont {D.}~\bibnamefont {Bruss}},\ }\href
  {https://doi.org/10.1002/qute.202000025} {\bibfield  {journal} {\bibinfo
  {journal} {Adv. Quant. Tech.}\ }\textbf {\bibinfo {volume} {14}},\ \bibinfo
  {pages} {2000025} (\bibinfo {year} {2020})}\BibitemShut {NoStop}%
\bibitem [{\citenamefont {Fitzsimons}\ and\ \citenamefont
  {Kashefi}(2017)}]{Fitzsimons:2017ge}%
  \BibitemOpen
  \bibfield  {author} {\bibinfo {author} {\bibfnamefont {J.~F.}\ \bibnamefont
  {Fitzsimons}}\ and\ \bibinfo {author} {\bibfnamefont {E.}~\bibnamefont
  {Kashefi}},\ }\href {https://doi.org/10.1103/PhysRevA.96.012303} {\bibfield
  {journal} {\bibinfo  {journal} {Phys. Rev. A}\ }\textbf {\bibinfo {volume}
  {96}},\ \bibinfo {pages} {012303} (\bibinfo {year} {2017})}\BibitemShut
  {NoStop}%
\bibitem [{\citenamefont {Briegel}\ \emph {et~al.}(1998)\citenamefont
  {Briegel}, \citenamefont {D{\"u}r}, \citenamefont {Cirac},\ and\
  \citenamefont {Zoller}}]{briegel1998quantum}%
  \BibitemOpen
  \bibfield  {author} {\bibinfo {author} {\bibfnamefont {H.-J.}\ \bibnamefont
  {Briegel}}, \bibinfo {author} {\bibfnamefont {W.}~\bibnamefont {D{\"u}r}},
  \bibinfo {author} {\bibfnamefont {J.~I.}\ \bibnamefont {Cirac}},\ and\
  \bibinfo {author} {\bibfnamefont {P.}~\bibnamefont {Zoller}},\ }\href
  {https://doi.org/10.1103/PhysRevLett.81.5932} {\bibfield  {journal} {\bibinfo
   {journal} {Phys. Rev. Lett.}\ }\textbf {\bibinfo {volume} {81}},\ \bibinfo
  {pages} {5932} (\bibinfo {year} {1998})}\BibitemShut {NoStop}%
\bibitem [{\citenamefont {D\"ur}\ \emph {et~al.}(1999)\citenamefont {D\"ur},
  \citenamefont {Briegel}, \citenamefont {Cirac},\ and\ \citenamefont
  {Zoller}}]{Duer1999repeater}%
  \BibitemOpen
  \bibfield  {author} {\bibinfo {author} {\bibfnamefont {W.}~\bibnamefont
  {D\"ur}}, \bibinfo {author} {\bibfnamefont {H.-J.}\ \bibnamefont {Briegel}},
  \bibinfo {author} {\bibfnamefont {J.~I.}\ \bibnamefont {Cirac}},\ and\
  \bibinfo {author} {\bibfnamefont {P.}~\bibnamefont {Zoller}},\ }\href
  {https://doi.org/10.1103/PhysRevA.59.169} {\bibfield  {journal} {\bibinfo
  {journal} {Phys. Rev. A}\ }\textbf {\bibinfo {volume} {59}},\ \bibinfo
  {pages} {169} (\bibinfo {year} {1999})}\BibitemShut {NoStop}%
\bibitem [{\citenamefont {Luong}\ \emph {et~al.}(2016)\citenamefont {Luong},
  \citenamefont {Jiang}, \citenamefont {Kim},\ and\ \citenamefont
  {L{\"u}tkenhaus}}]{Luong2016}%
  \BibitemOpen
  \bibfield  {author} {\bibinfo {author} {\bibfnamefont {D.}~\bibnamefont
  {Luong}}, \bibinfo {author} {\bibfnamefont {L.}~\bibnamefont {Jiang}},
  \bibinfo {author} {\bibfnamefont {J.}~\bibnamefont {Kim}},\ and\ \bibinfo
  {author} {\bibfnamefont {N.}~\bibnamefont {L{\"u}tkenhaus}},\ }\href
  {https://doi.org/10.1007/s00340-016-6373-4} {\bibfield  {journal} {\bibinfo
  {journal} {Appl. Phys. B}\ }\textbf {\bibinfo {volume} {122}},\ \bibinfo
  {pages} {96} (\bibinfo {year} {2016})}\BibitemShut {NoStop}%
\bibitem [{\citenamefont {Duan}\ \emph {et~al.}(2001)\citenamefont {Duan},
  \citenamefont {Lukin}, \citenamefont {Cirac},\ and\ \citenamefont
  {Zoller}}]{RepeaterLukin}%
  \BibitemOpen
  \bibfield  {author} {\bibinfo {author} {\bibfnamefont {L.-M.}\ \bibnamefont
  {Duan}}, \bibinfo {author} {\bibfnamefont {M.}~\bibnamefont {Lukin}},
  \bibinfo {author} {\bibfnamefont {J.~I.}\ \bibnamefont {Cirac}},\ and\
  \bibinfo {author} {\bibfnamefont {P.}~\bibnamefont {Zoller}},\ }\href
  {https://doi.org/10.1038/35106500} {\bibfield  {journal} {\bibinfo  {journal}
  {Nature}\ }\textbf {\bibinfo {volume} {414}},\ \bibinfo {pages} {413}
  (\bibinfo {year} {2001})}\BibitemShut {NoStop}%
\bibitem [{\citenamefont {Boaron}\ \emph {et~al.}(2018)\citenamefont {Boaron},
  \citenamefont {Boso}, \citenamefont {Rusca}, \citenamefont {Vulliez},
  \citenamefont {Autebert}, \citenamefont {Caloz}, \citenamefont {Perrenoud},
  \citenamefont {Gras}, \citenamefont {Bussi\`eres}, \citenamefont {Li},
  \citenamefont {Nolan}, \citenamefont {Martin},\ and\ \citenamefont
  {Zbinden}}]{PhysRevLett.121.190502}%
  \BibitemOpen
  \bibfield  {author} {\bibinfo {author} {\bibfnamefont {A.}~\bibnamefont
  {Boaron}}, \bibinfo {author} {\bibfnamefont {G.}~\bibnamefont {Boso}},
  \bibinfo {author} {\bibfnamefont {D.}~\bibnamefont {Rusca}}, \bibinfo
  {author} {\bibfnamefont {C.}~\bibnamefont {Vulliez}}, \bibinfo {author}
  {\bibfnamefont {C.}~\bibnamefont {Autebert}}, \bibinfo {author}
  {\bibfnamefont {M.}~\bibnamefont {Caloz}}, \bibinfo {author} {\bibfnamefont
  {M.}~\bibnamefont {Perrenoud}}, \bibinfo {author} {\bibfnamefont
  {G.}~\bibnamefont {Gras}}, \bibinfo {author} {\bibfnamefont {F.}~\bibnamefont
  {Bussi\`eres}}, \bibinfo {author} {\bibfnamefont {M.-J.}\ \bibnamefont {Li}},
  \bibinfo {author} {\bibfnamefont {D.}~\bibnamefont {Nolan}}, \bibinfo
  {author} {\bibfnamefont {A.}~\bibnamefont {Martin}},\ and\ \bibinfo {author}
  {\bibfnamefont {H.}~\bibnamefont {Zbinden}},\ }\href
  {https://doi.org/10.1103/PhysRevLett.121.190502} {\bibfield  {journal}
  {\bibinfo  {journal} {Phys. Rev. Lett.}\ }\textbf {\bibinfo {volume} {121}},\
  \bibinfo {pages} {190502} (\bibinfo {year} {2018})}\BibitemShut {NoStop}%
\bibitem [{\citenamefont {Yin}\ \emph {et~al.}(2016)\citenamefont {Yin},
  \citenamefont {Chen}, \citenamefont {Yu}, \citenamefont {Liu}, \citenamefont
  {You}, \citenamefont {Zhou}, \citenamefont {Chen}, \citenamefont {Mao},
  \citenamefont {Huang}, \citenamefont {Zhang}, \citenamefont {Chen},
  \citenamefont {Li}, \citenamefont {Nolan}, \citenamefont {Zhou},
  \citenamefont {Jiang}, \citenamefont {Wang}, \citenamefont {Zhang},
  \citenamefont {Wang},\ and\ \citenamefont {Pan}}]{PhysRevLett.117.190501}%
  \BibitemOpen
  \bibfield  {author} {\bibinfo {author} {\bibfnamefont {H.-L.}\ \bibnamefont
  {Yin}}, \bibinfo {author} {\bibfnamefont {T.-Y.}\ \bibnamefont {Chen}},
  \bibinfo {author} {\bibfnamefont {Z.-W.}\ \bibnamefont {Yu}}, \bibinfo
  {author} {\bibfnamefont {H.}~\bibnamefont {Liu}}, \bibinfo {author}
  {\bibfnamefont {L.-X.}\ \bibnamefont {You}}, \bibinfo {author} {\bibfnamefont
  {Y.-H.}\ \bibnamefont {Zhou}}, \bibinfo {author} {\bibfnamefont {S.-J.}\
  \bibnamefont {Chen}}, \bibinfo {author} {\bibfnamefont {Y.}~\bibnamefont
  {Mao}}, \bibinfo {author} {\bibfnamefont {M.-Q.}\ \bibnamefont {Huang}},
  \bibinfo {author} {\bibfnamefont {W.-J.}\ \bibnamefont {Zhang}}, \bibinfo
  {author} {\bibfnamefont {H.}~\bibnamefont {Chen}}, \bibinfo {author}
  {\bibfnamefont {M.~J.}\ \bibnamefont {Li}}, \bibinfo {author} {\bibfnamefont
  {D.}~\bibnamefont {Nolan}}, \bibinfo {author} {\bibfnamefont
  {F.}~\bibnamefont {Zhou}}, \bibinfo {author} {\bibfnamefont {X.}~\bibnamefont
  {Jiang}}, \bibinfo {author} {\bibfnamefont {Z.}~\bibnamefont {Wang}},
  \bibinfo {author} {\bibfnamefont {Q.}~\bibnamefont {Zhang}}, \bibinfo
  {author} {\bibfnamefont {X.-B.}\ \bibnamefont {Wang}},\ and\ \bibinfo
  {author} {\bibfnamefont {J.-W.}\ \bibnamefont {Pan}},\ }\href
  {https://doi.org/10.1103/PhysRevLett.117.190501} {\bibfield  {journal}
  {\bibinfo  {journal} {Phys. Rev. Lett.}\ }\textbf {\bibinfo {volume} {117}},\
  \bibinfo {pages} {190501} (\bibinfo {year} {2016})}\BibitemShut {NoStop}%
\bibitem [{\citenamefont {Chen}\ \emph {et~al.}(2021)\citenamefont {Chen},
  \citenamefont {Zhang}, \citenamefont {Chen}, \citenamefont {Cai},
  \citenamefont {Liao}, \citenamefont {Zhang}, \citenamefont {Chen},
  \citenamefont {Yin}, \citenamefont {Ren}, \citenamefont {Chen}, \citenamefont
  {Han}, \citenamefont {Yu}, \citenamefont {Liang}, \citenamefont {Zhou},
  \citenamefont {Yuan}, \citenamefont {Zhao}, \citenamefont {Wang},
  \citenamefont {Jiang}, \citenamefont {Zhang}, \citenamefont {Liu},
  \citenamefont {Li}, \citenamefont {Shen}, \citenamefont {Cao}, \citenamefont
  {Lu}, \citenamefont {Shu}, \citenamefont {Wang}, \citenamefont {Li},
  \citenamefont {Liu}, \citenamefont {Xu}, \citenamefont {Wang}, \citenamefont
  {Peng},\ and\ \citenamefont {Pan}}]{Chinese4600km}%
  \BibitemOpen
  \bibfield  {author} {\bibinfo {author} {\bibfnamefont {Y.-A.}\ \bibnamefont
  {Chen}}, \bibinfo {author} {\bibfnamefont {Q.}~\bibnamefont {Zhang}},
  \bibinfo {author} {\bibfnamefont {T.-Y.}\ \bibnamefont {Chen}}, \bibinfo
  {author} {\bibfnamefont {W.-Q.}\ \bibnamefont {Cai}}, \bibinfo {author}
  {\bibfnamefont {S.-K.}\ \bibnamefont {Liao}}, \bibinfo {author}
  {\bibfnamefont {J.}~\bibnamefont {Zhang}}, \bibinfo {author} {\bibfnamefont
  {K.}~\bibnamefont {Chen}}, \bibinfo {author} {\bibfnamefont {J.}~\bibnamefont
  {Yin}}, \bibinfo {author} {\bibfnamefont {J.-G.}\ \bibnamefont {Ren}},
  \bibinfo {author} {\bibfnamefont {Z.}~\bibnamefont {Chen}}, \bibinfo {author}
  {\bibfnamefont {S.-L.}\ \bibnamefont {Han}}, \bibinfo {author} {\bibfnamefont
  {Q.}~\bibnamefont {Yu}}, \bibinfo {author} {\bibfnamefont {K.}~\bibnamefont
  {Liang}}, \bibinfo {author} {\bibfnamefont {F.}~\bibnamefont {Zhou}},
  \bibinfo {author} {\bibfnamefont {X.}~\bibnamefont {Yuan}}, \bibinfo {author}
  {\bibfnamefont {M.-S.}\ \bibnamefont {Zhao}}, \bibinfo {author}
  {\bibfnamefont {T.-Y.}\ \bibnamefont {Wang}}, \bibinfo {author}
  {\bibfnamefont {X.}~\bibnamefont {Jiang}}, \bibinfo {author} {\bibfnamefont
  {L.}~\bibnamefont {Zhang}}, \bibinfo {author} {\bibfnamefont {W.-Y.}\
  \bibnamefont {Liu}}, \bibinfo {author} {\bibfnamefont {Y.}~\bibnamefont
  {Li}}, \bibinfo {author} {\bibfnamefont {Q.}~\bibnamefont {Shen}}, \bibinfo
  {author} {\bibfnamefont {Y.}~\bibnamefont {Cao}}, \bibinfo {author}
  {\bibfnamefont {C.-Y.}\ \bibnamefont {Lu}}, \bibinfo {author} {\bibfnamefont
  {R.}~\bibnamefont {Shu}}, \bibinfo {author} {\bibfnamefont {J.-Y.}\
  \bibnamefont {Wang}}, \bibinfo {author} {\bibfnamefont {L.}~\bibnamefont
  {Li}}, \bibinfo {author} {\bibfnamefont {N.-L.}\ \bibnamefont {Liu}},
  \bibinfo {author} {\bibfnamefont {F.}~\bibnamefont {Xu}}, \bibinfo {author}
  {\bibfnamefont {X.-B.}\ \bibnamefont {Wang}}, \bibinfo {author}
  {\bibfnamefont {C.-Z.}\ \bibnamefont {Peng}},\ and\ \bibinfo {author}
  {\bibfnamefont {J.-W.}\ \bibnamefont {Pan}},\ }\href
  {https://doi.org/10.1038/s41586-020-03093-8} {\bibfield  {journal} {\bibinfo
  {journal} {Nature}\ }\textbf {\bibinfo {volume} {589}},\ \bibinfo {pages}
  {214–219} (\bibinfo {year} {2021})}\BibitemShut {NoStop}%
\bibitem [{\citenamefont {Pirandola}\ \emph {et~al.}(2017)\citenamefont
  {Pirandola}, \citenamefont {Laurenza}, \citenamefont {Ottaviani},\ and\
  \citenamefont {Banchi}}]{PLOB}%
  \BibitemOpen
  \bibfield  {author} {\bibinfo {author} {\bibfnamefont {S.}~\bibnamefont
  {Pirandola}}, \bibinfo {author} {\bibfnamefont {R.}~\bibnamefont {Laurenza}},
  \bibinfo {author} {\bibfnamefont {C.}~\bibnamefont {Ottaviani}},\ and\
  \bibinfo {author} {\bibfnamefont {L.}~\bibnamefont {Banchi}},\ }\href
  {https://doi.org/10.1038/ncomms15043} {\bibfield  {journal} {\bibinfo
  {journal} {Nature Comm.}\ }\textbf {\bibinfo {volume} {8}},\ \bibinfo {pages}
  {15043} (\bibinfo {year} {2017})}\BibitemShut {NoStop}%
\bibitem [{\citenamefont {Wilde}\ \emph {et~al.}(2017)\citenamefont {Wilde},
  \citenamefont {Tomamichel},\ and\ \citenamefont {Berta}}]{Wilde2017}%
  \BibitemOpen
  \bibfield  {author} {\bibinfo {author} {\bibfnamefont {M.~M.}\ \bibnamefont
  {Wilde}}, \bibinfo {author} {\bibfnamefont {M.}~\bibnamefont {Tomamichel}},\
  and\ \bibinfo {author} {\bibfnamefont {M.}~\bibnamefont {Berta}},\ }\href
  {https://doi.org/10.1109/TIT.2017.2648825} {\bibfield  {journal} {\bibinfo
  {journal} {IEEE Trans. Inf. Th.}\ }\textbf {\bibinfo {volume} {63}},\
  \bibinfo {pages} {1792–1817} (\bibinfo {year} {2017})}\BibitemShut
  {NoStop}%
\bibitem [{\citenamefont {Christandl}\ and\ \citenamefont
  {M{\"u}ller-Hermes}(2017)}]{Christandl2017}%
  \BibitemOpen
  \bibfield  {author} {\bibinfo {author} {\bibfnamefont {M.}~\bibnamefont
  {Christandl}}\ and\ \bibinfo {author} {\bibfnamefont {A.}~\bibnamefont
  {M{\"u}ller-Hermes}},\ }\href {https://doi.org/10.1007/s00220-017-2885-y}
  {\bibfield  {journal} {\bibinfo  {journal} {Communications in Mathematical
  Physics}\ }\textbf {\bibinfo {volume} {353}},\ \bibinfo {pages} {821}
  (\bibinfo {year} {2017})}\BibitemShut {NoStop}%
\bibitem [{\citenamefont {Laurenza}\ \emph {et~al.}(2022)\citenamefont
  {Laurenza}, \citenamefont {Walk}, \citenamefont {Eisert},\ and\ \citenamefont
  {Pirandola}}]{Laurenza}%
  \BibitemOpen
  \bibfield  {author} {\bibinfo {author} {\bibfnamefont {R.}~\bibnamefont
  {Laurenza}}, \bibinfo {author} {\bibfnamefont {N.}~\bibnamefont {Walk}},
  \bibinfo {author} {\bibfnamefont {J.}~\bibnamefont {Eisert}},\ and\ \bibinfo
  {author} {\bibfnamefont {S.}~\bibnamefont {Pirandola}},\ }\href
  {https://doi.org/10.1103/PhysRevResearch.4.023158} {\bibfield  {journal}
  {\bibinfo  {journal} {Phys. Rev. Research}\ }\textbf {\bibinfo {volume}
  {4}},\ \bibinfo {pages} {023158} (\bibinfo {year} {2022})}\BibitemShut
  {NoStop}%
\bibitem [{\citenamefont {Harney}\ and\ \citenamefont
  {Pirandola}(2022{\natexlab{a}})}]{PirandolaRepeaters}%
  \BibitemOpen
  \bibfield  {author} {\bibinfo {author} {\bibfnamefont {C.}~\bibnamefont
  {Harney}}\ and\ \bibinfo {author} {\bibfnamefont {S.}~\bibnamefont
  {Pirandola}},\ }\href {https://doi.org/10.1088/2058-9565/ac7ba0} {\bibfield
  {journal} {\bibinfo  {journal} {Quantum Sci. Technol.}\ }\textbf {\bibinfo
  {volume} {7}},\ \bibinfo {pages} {045009} (\bibinfo {year}
  {2022}{\natexlab{a}})}\BibitemShut {NoStop}%
\bibitem [{\citenamefont {Muralidharan}\ \emph {et~al.}(2016)\citenamefont
  {Muralidharan}, \citenamefont {Li}, \citenamefont {Kim}, \citenamefont
  {Lütkenhaus}, \citenamefont {Lukin},\ and\ \citenamefont
  {Jiang}}]{Challenge1}%
  \BibitemOpen
  \bibfield  {author} {\bibinfo {author} {\bibfnamefont {S.}~\bibnamefont
  {Muralidharan}}, \bibinfo {author} {\bibfnamefont {L.}~\bibnamefont {Li}},
  \bibinfo {author} {\bibfnamefont {J.}~\bibnamefont {Kim}}, \bibinfo {author}
  {\bibfnamefont {N.}~\bibnamefont {Lütkenhaus}}, \bibinfo {author}
  {\bibfnamefont {M.~D.}\ \bibnamefont {Lukin}},\ and\ \bibinfo {author}
  {\bibfnamefont {L.}~\bibnamefont {Jiang}},\ }\href
  {https://doi.org/10.1038/srep20463} {\bibfield  {journal} {\bibinfo
  {journal} {Sci. Rep.}\ }\textbf {\bibinfo {volume} {6}},\ \bibinfo {pages}
  {20463} (\bibinfo {year} {2016})}\BibitemShut {NoStop}%
\bibitem [{\citenamefont {Harney}\ and\ \citenamefont
  {Pirandola}(2022{\natexlab{b}})}]{PRXQuantum.3.010349}%
  \BibitemOpen
  \bibfield  {author} {\bibinfo {author} {\bibfnamefont {C.}~\bibnamefont
  {Harney}}\ and\ \bibinfo {author} {\bibfnamefont {S.}~\bibnamefont
  {Pirandola}},\ }\href {https://doi.org/10.1103/PRXQuantum.3.010349}
  {\bibfield  {journal} {\bibinfo  {journal} {PRX Quantum}\ }\textbf {\bibinfo
  {volume} {3}},\ \bibinfo {pages} {010349} (\bibinfo {year}
  {2022}{\natexlab{b}})}\BibitemShut {NoStop}%
\bibitem [{\citenamefont {van Loock}\ \emph {et~al.}(2020)\citenamefont {van
  Loock}, \citenamefont {Alt}, \citenamefont {Becher}, \citenamefont {Benson},
  \citenamefont {Boche}, \citenamefont {Deppe}, \citenamefont {Eschner},
  \citenamefont {Höfling}, \citenamefont {Meschede}, \citenamefont {Michler},
  \citenamefont {Schmidt},\ and\ \citenamefont
  {Weinfurter}}]{ExtendingQuantumLinks}%
  \BibitemOpen
  \bibfield  {author} {\bibinfo {author} {\bibfnamefont {P.}~\bibnamefont {van
  Loock}}, \bibinfo {author} {\bibfnamefont {W.}~\bibnamefont {Alt}}, \bibinfo
  {author} {\bibfnamefont {C.}~\bibnamefont {Becher}}, \bibinfo {author}
  {\bibfnamefont {O.}~\bibnamefont {Benson}}, \bibinfo {author} {\bibfnamefont
  {H.}~\bibnamefont {Boche}}, \bibinfo {author} {\bibfnamefont
  {C.}~\bibnamefont {Deppe}}, \bibinfo {author} {\bibfnamefont
  {J.}~\bibnamefont {Eschner}}, \bibinfo {author} {\bibfnamefont
  {S.}~\bibnamefont {Höfling}}, \bibinfo {author} {\bibfnamefont
  {D.}~\bibnamefont {Meschede}}, \bibinfo {author} {\bibfnamefont
  {P.}~\bibnamefont {Michler}}, \bibinfo {author} {\bibfnamefont
  {F.}~\bibnamefont {Schmidt}},\ and\ \bibinfo {author} {\bibfnamefont
  {H.}~\bibnamefont {Weinfurter}},\ }\href
  {https://doi.org/10.1002/qute.201900141} {\bibfield  {journal} {\bibinfo
  {journal} {Adv. Quant. Tech.}\ }\textbf {\bibinfo {volume} {3}},\ \bibinfo
  {pages} {1900141} (\bibinfo {year} {2020})}\BibitemShut {NoStop}%
\bibitem [{\citenamefont {Shchukin}\ \emph {et~al.}(2019)\citenamefont
  {Shchukin}, \citenamefont {Schmidt},\ and\ \citenamefont {van
  Loock}}]{Shchukin2019}%
  \BibitemOpen
  \bibfield  {author} {\bibinfo {author} {\bibfnamefont {E.}~\bibnamefont
  {Shchukin}}, \bibinfo {author} {\bibfnamefont {F.}~\bibnamefont {Schmidt}},\
  and\ \bibinfo {author} {\bibfnamefont {P.}~\bibnamefont {van Loock}},\ }\href
  {https://doi.org/10.1103/PhysRevA.100.032322} {\bibfield  {journal} {\bibinfo
   {journal} {Phys. Rev. A}\ }\textbf {\bibinfo {volume} {100}},\ \bibinfo
  {pages} {032322} (\bibinfo {year} {2019})}\BibitemShut {NoStop}%
\bibitem [{\citenamefont {Tr\'enyi}\ and\ \citenamefont
  {L\"utkenhaus}(2020)}]{Trenyi2020}%
  \BibitemOpen
  \bibfield  {author} {\bibinfo {author} {\bibfnamefont {R.}~\bibnamefont
  {Tr\'enyi}}\ and\ \bibinfo {author} {\bibfnamefont {N.}~\bibnamefont
  {L\"utkenhaus}},\ }\href {https://doi.org/10.1103/PhysRevA.101.012325}
  {\bibfield  {journal} {\bibinfo  {journal} {Phys. Rev. A}\ }\textbf {\bibinfo
  {volume} {101}},\ \bibinfo {pages} {012325} (\bibinfo {year}
  {2020})}\BibitemShut {NoStop}%
\bibitem [{\citenamefont {Kamin}\ \emph {et~al.}(2023)\citenamefont {Kamin},
  \citenamefont {Shchukin}, \citenamefont {Schmidt},\ and\ \citenamefont {van
  Loock}}]{Kamin2022}%
  \BibitemOpen
  \bibfield  {author} {\bibinfo {author} {\bibfnamefont {L.}~\bibnamefont
  {Kamin}}, \bibinfo {author} {\bibfnamefont {E.}~\bibnamefont {Shchukin}},
  \bibinfo {author} {\bibfnamefont {F.}~\bibnamefont {Schmidt}},\ and\ \bibinfo
  {author} {\bibfnamefont {P.}~\bibnamefont {van Loock}},\ }\href
  {https://doi.org/10.1103/PhysRevResearch.5.023086} {\bibfield  {journal}
  {\bibinfo  {journal} {Phys. Rev. Res.}\ }\textbf {\bibinfo {volume} {5}},\
  \bibinfo {pages} {023086} (\bibinfo {year} {2023})}\BibitemShut {NoStop}%
\bibitem [{\citenamefont {Coopmans}\ \emph {et~al.}(2021)\citenamefont
  {Coopmans}, \citenamefont {Knegjens}, \citenamefont {Dahlberg}, \citenamefont
  {Maier}, \citenamefont {Nijsten}, \citenamefont {de~Oliveira~Filho},
  \citenamefont {Papendrecht}, \citenamefont {Rabbie}, \citenamefont
  {Rozpędek}, \citenamefont {Skrzypczyk}, \citenamefont {Wubben},
  \citenamefont {de~Jong}, \citenamefont {Podareanu}, \citenamefont
  {Torres-Knoop}, \citenamefont {Elkouss},\ and\ \citenamefont
  {Wehner}}]{NetSquid}%
  \BibitemOpen
  \bibfield  {author} {\bibinfo {author} {\bibfnamefont {T.}~\bibnamefont
  {Coopmans}}, \bibinfo {author} {\bibfnamefont {R.}~\bibnamefont {Knegjens}},
  \bibinfo {author} {\bibfnamefont {A.}~\bibnamefont {Dahlberg}}, \bibinfo
  {author} {\bibfnamefont {D.}~\bibnamefont {Maier}}, \bibinfo {author}
  {\bibfnamefont {L.}~\bibnamefont {Nijsten}}, \bibinfo {author} {\bibfnamefont
  {J.}~\bibnamefont {de~Oliveira~Filho}}, \bibinfo {author} {\bibfnamefont
  {M.}~\bibnamefont {Papendrecht}}, \bibinfo {author} {\bibfnamefont
  {J.}~\bibnamefont {Rabbie}}, \bibinfo {author} {\bibfnamefont
  {F.}~\bibnamefont {Rozpędek}}, \bibinfo {author} {\bibfnamefont
  {M.}~\bibnamefont {Skrzypczyk}}, \bibinfo {author} {\bibfnamefont
  {L.}~\bibnamefont {Wubben}}, \bibinfo {author} {\bibfnamefont
  {W.}~\bibnamefont {de~Jong}}, \bibinfo {author} {\bibfnamefont
  {D.}~\bibnamefont {Podareanu}}, \bibinfo {author} {\bibfnamefont
  {A.}~\bibnamefont {Torres-Knoop}}, \bibinfo {author} {\bibfnamefont
  {D.}~\bibnamefont {Elkouss}},\ and\ \bibinfo {author} {\bibfnamefont
  {S.}~\bibnamefont {Wehner}},\ }\href
  {https://doi.org/10.1038/s42005-021-00647-8} {\bibfield  {journal} {\bibinfo
  {journal} {Commun. Phys.}\ }\textbf {\bibinfo {volume} {4}},\ \bibinfo
  {pages} {164} (\bibinfo {year} {2021})}\BibitemShut {NoStop}%
\bibitem [{\citenamefont {Satoh}\ \emph {et~al.}(2022)\citenamefont {Satoh},
  \citenamefont {Hajdušek}, \citenamefont {Benchasattabuse}, \citenamefont
  {Nagayama}, \citenamefont {Teramoto}, \citenamefont {Matsuo}, \citenamefont
  {Metwalli}, \citenamefont {Pathumsoot}, \citenamefont {Satoh}, \citenamefont
  {Suzuki},\ and\ \citenamefont {Meter}}]{satoh2021quisp}%
  \BibitemOpen
  \bibfield  {author} {\bibinfo {author} {\bibfnamefont {R.}~\bibnamefont
  {Satoh}}, \bibinfo {author} {\bibfnamefont {M.}~\bibnamefont {Hajdušek}},
  \bibinfo {author} {\bibfnamefont {N.}~\bibnamefont {Benchasattabuse}},
  \bibinfo {author} {\bibfnamefont {S.}~\bibnamefont {Nagayama}}, \bibinfo
  {author} {\bibfnamefont {K.}~\bibnamefont {Teramoto}}, \bibinfo {author}
  {\bibfnamefont {T.}~\bibnamefont {Matsuo}}, \bibinfo {author} {\bibfnamefont
  {S.~A.}\ \bibnamefont {Metwalli}}, \bibinfo {author} {\bibfnamefont
  {P.}~\bibnamefont {Pathumsoot}}, \bibinfo {author} {\bibfnamefont
  {T.}~\bibnamefont {Satoh}}, \bibinfo {author} {\bibfnamefont
  {S.}~\bibnamefont {Suzuki}},\ and\ \bibinfo {author} {\bibfnamefont {R.~V.}\
  \bibnamefont {Meter}},\ }in\ \href
  {https://doi.org/10.1109/QCE53715.2022.00056} {\emph {\bibinfo {booktitle}
  {2022 IEEE International Conference on Quantum Computing and Engineering
  (QCE)}}}\ (\bibinfo {year} {2022})\ pp.\ \bibinfo {pages}
  {353--364}\BibitemShut {NoStop}%
\bibitem [{\citenamefont {Wu}\ \emph {et~al.}(2021{\natexlab{a}})\citenamefont
  {Wu}, \citenamefont {Kolar}, \citenamefont {Chung}, \citenamefont {Jin},
  \citenamefont {Zhong}, \citenamefont {Kettimuthu},\ and\ \citenamefont
  {Suchara}}]{sequence}%
  \BibitemOpen
  \bibfield  {author} {\bibinfo {author} {\bibfnamefont {X.}~\bibnamefont
  {Wu}}, \bibinfo {author} {\bibfnamefont {A.}~\bibnamefont {Kolar}}, \bibinfo
  {author} {\bibfnamefont {J.}~\bibnamefont {Chung}}, \bibinfo {author}
  {\bibfnamefont {D.}~\bibnamefont {Jin}}, \bibinfo {author} {\bibfnamefont
  {T.}~\bibnamefont {Zhong}}, \bibinfo {author} {\bibfnamefont
  {R.}~\bibnamefont {Kettimuthu}},\ and\ \bibinfo {author} {\bibfnamefont
  {M.}~\bibnamefont {Suchara}},\ }\href
  {https://doi.org/10.1088/2058-9565/ac22f6} {\bibfield  {journal} {\bibinfo
  {journal} {Quant Sc. Tech.}\ }\textbf {\bibinfo {volume} {6}},\ \bibinfo
  {pages} {045027} (\bibinfo {year} {2021}{\natexlab{a}})}\BibitemShut
  {NoStop}%
\bibitem [{\citenamefont {Avis}\ \emph {et~al.}(2023)\citenamefont {Avis},
  \citenamefont {Ferreira~da Silva}, \citenamefont {Coopmans}, \citenamefont
  {Dahlberg}, \citenamefont {Jirovsk{\'a}}, \citenamefont {Maier},
  \citenamefont {Rabbie}, \citenamefont {Torres-Knoop},\ and\ \citenamefont
  {Wehner}}]{Avis22}%
  \BibitemOpen
  \bibfield  {author} {\bibinfo {author} {\bibfnamefont {G.}~\bibnamefont
  {Avis}}, \bibinfo {author} {\bibfnamefont {F.}~\bibnamefont {Ferreira~da
  Silva}}, \bibinfo {author} {\bibfnamefont {T.}~\bibnamefont {Coopmans}},
  \bibinfo {author} {\bibfnamefont {A.}~\bibnamefont {Dahlberg}}, \bibinfo
  {author} {\bibfnamefont {H.}~\bibnamefont {Jirovsk{\'a}}}, \bibinfo {author}
  {\bibfnamefont {D.}~\bibnamefont {Maier}}, \bibinfo {author} {\bibfnamefont
  {J.}~\bibnamefont {Rabbie}}, \bibinfo {author} {\bibfnamefont
  {A.}~\bibnamefont {Torres-Knoop}},\ and\ \bibinfo {author} {\bibfnamefont
  {S.}~\bibnamefont {Wehner}},\ }\href
  {https://doi.org/10.1038/s41534-023-00765-x} {\bibfield  {journal} {\bibinfo
  {journal} {npj Quantum Information}\ }\textbf {\bibinfo {volume} {9}},\
  \bibinfo {pages} {100} (\bibinfo {year} {2023})}\BibitemShut {NoStop}%
\bibitem [{\citenamefont {da~Silva}\ \emph {et~al.}(2021)\citenamefont
  {da~Silva}, \citenamefont {Torres-Knoop}, \citenamefont {Coopmans},
  \citenamefont {Maier},\ and\ \citenamefont
  {Wehner}}]{Ferreira_da_Silva_2021}%
  \BibitemOpen
  \bibfield  {author} {\bibinfo {author} {\bibfnamefont {F.~F.}\ \bibnamefont
  {da~Silva}}, \bibinfo {author} {\bibfnamefont {A.}~\bibnamefont
  {Torres-Knoop}}, \bibinfo {author} {\bibfnamefont {T.}~\bibnamefont
  {Coopmans}}, \bibinfo {author} {\bibfnamefont {D.}~\bibnamefont {Maier}},\
  and\ \bibinfo {author} {\bibfnamefont {S.}~\bibnamefont {Wehner}},\ }\href
  {https://doi.org/10.1088/2058-9565/abfc93} {\bibfield  {journal} {\bibinfo
  {journal} {Quant. Sc. Tech.}\ }\textbf {\bibinfo {volume} {6}},\ \bibinfo
  {pages} {035007} (\bibinfo {year} {2021})}\BibitemShut {NoStop}%
\bibitem [{\citenamefont {Wallnöfer}(2022)}]{requsim}%
  \BibitemOpen
  \bibfield  {author} {\bibinfo {author} {\bibfnamefont {J.}~\bibnamefont
  {Wallnöfer}},\ }\href@noop {} {\bibinfo {title} {{R}e{Q}u{S}im}},\ \bibinfo
  {howpublished}
  {\href{https://doi.org/10.5281/zenodo.7290708}{10.5281/zenodo.7290708}}
  (\bibinfo {year} {2022})\BibitemShut {NoStop}%
\bibitem [{\citenamefont {Shchukin}\ and\ \citenamefont {van
  Loock}(2022)}]{Shchukin2022}%
  \BibitemOpen
  \bibfield  {author} {\bibinfo {author} {\bibfnamefont {E.}~\bibnamefont
  {Shchukin}}\ and\ \bibinfo {author} {\bibfnamefont {P.}~\bibnamefont {van
  Loock}},\ }\href {https://doi.org/10.1103/PhysRevLett.128.150502} {\bibfield
  {journal} {\bibinfo  {journal} {Phys. Rev. Lett.}\ }\textbf {\bibinfo
  {volume} {128}},\ \bibinfo {pages} {150502} (\bibinfo {year}
  {2022})}\BibitemShut {NoStop}%
\bibitem [{\citenamefont {Zwerger}\ \emph {et~al.}(2016)\citenamefont
  {Zwerger}, \citenamefont {Briegel},\ and\ \citenamefont
  {D{\"u}r}}]{Zwerger2016}%
  \BibitemOpen
  \bibfield  {author} {\bibinfo {author} {\bibfnamefont {M.}~\bibnamefont
  {Zwerger}}, \bibinfo {author} {\bibfnamefont {H.~J.}\ \bibnamefont
  {Briegel}},\ and\ \bibinfo {author} {\bibfnamefont {W.}~\bibnamefont
  {D{\"u}r}},\ }\href {https://doi.org/10.1007/s00340-015-6285-8} {\bibfield
  {journal} {\bibinfo  {journal} {Appl. Phys. B}\ }\textbf {\bibinfo {volume}
  {122}},\ \bibinfo {pages} {50} (\bibinfo {year} {2016})}\BibitemShut
  {NoStop}%
\bibitem [{\citenamefont {Walln{\"o}fer}\ \emph {et~al.}(2019)\citenamefont
  {Walln{\"o}fer}, \citenamefont {Pirker}, \citenamefont {Zwerger},\ and\
  \citenamefont {D{\"u}r}}]{Wallnoefer2019}%
  \BibitemOpen
  \bibfield  {author} {\bibinfo {author} {\bibfnamefont {J.}~\bibnamefont
  {Walln{\"o}fer}}, \bibinfo {author} {\bibfnamefont {A.}~\bibnamefont
  {Pirker}}, \bibinfo {author} {\bibfnamefont {M.}~\bibnamefont {Zwerger}},\
  and\ \bibinfo {author} {\bibfnamefont {W.}~\bibnamefont {D{\"u}r}},\ }\href
  {https://doi.org/10.1038/s41598-018-36543-5} {\bibfield  {journal} {\bibinfo
  {journal} {Sci. Rep.}\ }\textbf {\bibinfo {volume} {9}},\ \bibinfo {pages}
  {314} (\bibinfo {year} {2019})}\BibitemShut {NoStop}%
\bibitem [{\citenamefont {Collins}\ \emph {et~al.}(2007)\citenamefont
  {Collins}, \citenamefont {Jenkins}, \citenamefont {Kuzmich},\ and\
  \citenamefont {Kennedy}}]{Collins2007}%
  \BibitemOpen
  \bibfield  {author} {\bibinfo {author} {\bibfnamefont {O.~A.}\ \bibnamefont
  {Collins}}, \bibinfo {author} {\bibfnamefont {S.~D.}\ \bibnamefont
  {Jenkins}}, \bibinfo {author} {\bibfnamefont {A.}~\bibnamefont {Kuzmich}},\
  and\ \bibinfo {author} {\bibfnamefont {T.~A.~B.}\ \bibnamefont {Kennedy}},\
  }\href {https://doi.org/10.1103/PhysRevLett.98.060502} {\bibfield  {journal}
  {\bibinfo  {journal} {Phys. Rev. Lett.}\ }\textbf {\bibinfo {volume} {98}},\
  \bibinfo {pages} {060502} (\bibinfo {year} {2007})}\BibitemShut {NoStop}%
\bibitem [{\citenamefont {Bernardes}\ \emph {et~al.}(2011)\citenamefont
  {Bernardes}, \citenamefont {Praxmeyer},\ and\ \citenamefont {van
  Loock}}]{Bernardes2011}%
  \BibitemOpen
  \bibfield  {author} {\bibinfo {author} {\bibfnamefont {N.~K.}\ \bibnamefont
  {Bernardes}}, \bibinfo {author} {\bibfnamefont {L.}~\bibnamefont
  {Praxmeyer}},\ and\ \bibinfo {author} {\bibfnamefont {P.}~\bibnamefont {van
  Loock}},\ }\href {https://doi.org/10.1103/PhysRevA.83.012323} {\bibfield
  {journal} {\bibinfo  {journal} {Phys. Rev. A}\ }\textbf {\bibinfo {volume}
  {83}},\ \bibinfo {pages} {012323} (\bibinfo {year} {2011})}\BibitemShut
  {NoStop}%
\bibitem [{\citenamefont {Deutsch}\ \emph {et~al.}(1996)\citenamefont
  {Deutsch}, \citenamefont {Ekert}, \citenamefont {Jozsa}, \citenamefont
  {Macchiavello}, \citenamefont {Popescu},\ and\ \citenamefont
  {Sanpera}}]{dejmps}%
  \BibitemOpen
  \bibfield  {author} {\bibinfo {author} {\bibfnamefont {D.}~\bibnamefont
  {Deutsch}}, \bibinfo {author} {\bibfnamefont {A.}~\bibnamefont {Ekert}},
  \bibinfo {author} {\bibfnamefont {R.}~\bibnamefont {Jozsa}}, \bibinfo
  {author} {\bibfnamefont {C.}~\bibnamefont {Macchiavello}}, \bibinfo {author}
  {\bibfnamefont {S.}~\bibnamefont {Popescu}},\ and\ \bibinfo {author}
  {\bibfnamefont {A.}~\bibnamefont {Sanpera}},\ }\href
  {https://doi.org/10.1103/PhysRevLett.77.2818} {\bibfield  {journal} {\bibinfo
   {journal} {Phys. Rev. Lett.}\ }\textbf {\bibinfo {volume} {77}},\ \bibinfo
  {pages} {2818} (\bibinfo {year} {1996})}\BibitemShut {NoStop}%
\bibitem [{\citenamefont {D\"ur}\ and\ \citenamefont
  {Briegel}(2003)}]{Duer2003}%
  \BibitemOpen
  \bibfield  {author} {\bibinfo {author} {\bibfnamefont {W.}~\bibnamefont
  {D\"ur}}\ and\ \bibinfo {author} {\bibfnamefont {H.-J.}\ \bibnamefont
  {Briegel}},\ }\href {https://doi.org/10.1103/PhysRevLett.90.067901}
  {\bibfield  {journal} {\bibinfo  {journal} {Phys. Rev. Lett.}\ }\textbf
  {\bibinfo {volume} {90}},\ \bibinfo {pages} {067901} (\bibinfo {year}
  {2003})}\BibitemShut {NoStop}%
\bibitem [{\citenamefont {Rozpędek}\ \emph {et~al.}(2018)\citenamefont
  {Rozpędek}, \citenamefont {Goodenough}, \citenamefont {Ribeiro},
  \citenamefont {Kalb}, \citenamefont {Vivoli}, \citenamefont {Reiserer},
  \citenamefont {Hanson}, \citenamefont {Wehner},\ and\ \citenamefont
  {Elkouss}}]{Rozpedek2018}%
  \BibitemOpen
  \bibfield  {author} {\bibinfo {author} {\bibfnamefont {F.}~\bibnamefont
  {Rozpędek}}, \bibinfo {author} {\bibfnamefont {K.}~\bibnamefont
  {Goodenough}}, \bibinfo {author} {\bibfnamefont {J.}~\bibnamefont {Ribeiro}},
  \bibinfo {author} {\bibfnamefont {N.}~\bibnamefont {Kalb}}, \bibinfo {author}
  {\bibfnamefont {V.~C.}\ \bibnamefont {Vivoli}}, \bibinfo {author}
  {\bibfnamefont {A.}~\bibnamefont {Reiserer}}, \bibinfo {author}
  {\bibfnamefont {R.}~\bibnamefont {Hanson}}, \bibinfo {author} {\bibfnamefont
  {S.}~\bibnamefont {Wehner}},\ and\ \bibinfo {author} {\bibfnamefont
  {D.}~\bibnamefont {Elkouss}},\ }\href
  {https://doi.org/10.1088/2058-9565/aab31b} {\bibfield  {journal} {\bibinfo
  {journal} {Quant. Sc. Technol.}\ }\textbf {\bibinfo {volume} {3}},\ \bibinfo
  {pages} {034002} (\bibinfo {year} {2018})}\BibitemShut {NoStop}%
\bibitem [{\citenamefont {Khatri}(2021)}]{Khatri2021}%
  \BibitemOpen
  \bibfield  {author} {\bibinfo {author} {\bibfnamefont {S.}~\bibnamefont
  {Khatri}},\ }\href {https://doi.org/10.22331/q-2021-09-07-537} {\bibfield
  {journal} {\bibinfo  {journal} {{Quantum}}\ }\textbf {\bibinfo {volume}
  {5}},\ \bibinfo {pages} {537} (\bibinfo {year} {2021})}\BibitemShut {NoStop}%
\bibitem [{\citenamefont {Khatri}(2022)}]{Khatri2022}%
  \BibitemOpen
  \bibfield  {author} {\bibinfo {author} {\bibfnamefont {S.}~\bibnamefont
  {Khatri}},\ }\href {https://doi.org/10.1116/5.0084653} {\bibfield  {journal}
  {\bibinfo  {journal} {AVS Quantum Science}\ }\textbf {\bibinfo {volume}
  {4}},\ \bibinfo {pages} {030501} (\bibinfo {year} {2022})}\BibitemShut
  {NoStop}%
\bibitem [{\citenamefont {Lucamarini}\ \emph {et~al.}(2018)\citenamefont
  {Lucamarini}, \citenamefont {Yuan}, \citenamefont {Dynes},\ and\
  \citenamefont {Shields}}]{Lucamarini2018}%
  \BibitemOpen
  \bibfield  {author} {\bibinfo {author} {\bibfnamefont {M.}~\bibnamefont
  {Lucamarini}}, \bibinfo {author} {\bibfnamefont {Z.~L.}\ \bibnamefont
  {Yuan}}, \bibinfo {author} {\bibfnamefont {J.~F.}\ \bibnamefont {Dynes}},\
  and\ \bibinfo {author} {\bibfnamefont {A.~J.}\ \bibnamefont {Shields}},\
  }\href {https://doi.org/10.1038/s41586-018-0066-6} {\bibfield  {journal}
  {\bibinfo  {journal} {Nature}\ }\textbf {\bibinfo {volume} {557}},\ \bibinfo
  {pages} {400} (\bibinfo {year} {2018})}\BibitemShut {NoStop}%
\bibitem [{\citenamefont {I{\~{n}}esta}\ \emph {et~al.}(2023)\citenamefont
  {I{\~{n}}esta}, \citenamefont {Vardoyan}, \citenamefont {Scavuzzo},\ and\
  \citenamefont {Wehner}}]{Inesta22}%
  \BibitemOpen
  \bibfield  {author} {\bibinfo {author} {\bibfnamefont {{\'A}.~G.}\
  \bibnamefont {I{\~{n}}esta}}, \bibinfo {author} {\bibfnamefont
  {G.}~\bibnamefont {Vardoyan}}, \bibinfo {author} {\bibfnamefont
  {L.}~\bibnamefont {Scavuzzo}},\ and\ \bibinfo {author} {\bibfnamefont
  {S.}~\bibnamefont {Wehner}},\ }\href
  {https://doi.org/10.1038/s41534-023-00713-9} {\bibfield  {journal} {\bibinfo
  {journal} {npj Quantum Information}\ }\textbf {\bibinfo {volume} {9}},\
  \bibinfo {pages} {46} (\bibinfo {year} {2023})}\BibitemShut {NoStop}%
\bibitem [{\citenamefont {G{\"u}ndo{\u{g}}an}\ \emph
  {et~al.}(2021)\citenamefont {G{\"u}ndo{\u{g}}an}, \citenamefont {Sidhu},
  \citenamefont {Henderson}, \citenamefont {Mazzarella}, \citenamefont
  {Wolters}, \citenamefont {Oi},\ and\ \citenamefont {Krutzik}}]{Gundogan2021}%
  \BibitemOpen
  \bibfield  {author} {\bibinfo {author} {\bibfnamefont {M.}~\bibnamefont
  {G{\"u}ndo{\u{g}}an}}, \bibinfo {author} {\bibfnamefont {J.~S.}\ \bibnamefont
  {Sidhu}}, \bibinfo {author} {\bibfnamefont {V.}~\bibnamefont {Henderson}},
  \bibinfo {author} {\bibfnamefont {L.}~\bibnamefont {Mazzarella}}, \bibinfo
  {author} {\bibfnamefont {J.}~\bibnamefont {Wolters}}, \bibinfo {author}
  {\bibfnamefont {D.~K.~L.}\ \bibnamefont {Oi}},\ and\ \bibinfo {author}
  {\bibfnamefont {M.}~\bibnamefont {Krutzik}},\ }\href
  {https://doi.org/10.1038/s41534-021-00460-9} {\bibfield  {journal} {\bibinfo
  {journal} {npj Quant. Inf.}\ }\textbf {\bibinfo {volume} {7}},\ \bibinfo
  {pages} {128} (\bibinfo {year} {2021})}\BibitemShut {NoStop}%
\bibitem [{\citenamefont {Walln\"ofer}\ \emph {et~al.}(2020)\citenamefont
  {Walln\"ofer}, \citenamefont {Melnikov}, \citenamefont {D\"ur},\ and\
  \citenamefont {Briegel}}]{Wallnoefer2020}%
  \BibitemOpen
  \bibfield  {author} {\bibinfo {author} {\bibfnamefont {J.}~\bibnamefont
  {Walln\"ofer}}, \bibinfo {author} {\bibfnamefont {A.~A.}\ \bibnamefont
  {Melnikov}}, \bibinfo {author} {\bibfnamefont {W.}~\bibnamefont {D\"ur}},\
  and\ \bibinfo {author} {\bibfnamefont {H.~J.}\ \bibnamefont {Briegel}},\
  }\href {https://doi.org/10.1103/PRXQuantum.1.010301} {\bibfield  {journal}
  {\bibinfo  {journal} {PRX Quantum}\ }\textbf {\bibinfo {volume} {1}},\
  \bibinfo {pages} {010301} (\bibinfo {year} {2020})}\BibitemShut {NoStop}%
\bibitem [{\citenamefont {Dahlberg}\ and\ \citenamefont
  {Wehner}(2018)}]{simulaqron}%
  \BibitemOpen
  \bibfield  {author} {\bibinfo {author} {\bibfnamefont {A.}~\bibnamefont
  {Dahlberg}}\ and\ \bibinfo {author} {\bibfnamefont {S.}~\bibnamefont
  {Wehner}},\ }\href {https://doi.org/10.1088/2058-9565/aad56e} {\bibfield
  {journal} {\bibinfo  {journal} {Quant. Sc. Technol.}\ }\textbf {\bibinfo
  {volume} {4}},\ \bibinfo {pages} {015001} (\bibinfo {year}
  {2018})}\BibitemShut {NoStop}%
\bibitem [{\citenamefont {Dahlberg}\ \emph {et~al.}(2022)\citenamefont
  {Dahlberg}, \citenamefont {van~der Vecht}, \citenamefont {Donne},
  \citenamefont {Skrzypczyk}, \citenamefont {te~Raa}, \citenamefont
  {Kozlowski},\ and\ \citenamefont {Wehner}}]{NetQASM}%
  \BibitemOpen
  \bibfield  {author} {\bibinfo {author} {\bibfnamefont {A.}~\bibnamefont
  {Dahlberg}}, \bibinfo {author} {\bibfnamefont {B.}~\bibnamefont {van~der
  Vecht}}, \bibinfo {author} {\bibfnamefont {C.~D.}\ \bibnamefont {Donne}},
  \bibinfo {author} {\bibfnamefont {M.}~\bibnamefont {Skrzypczyk}}, \bibinfo
  {author} {\bibfnamefont {I.}~\bibnamefont {te~Raa}}, \bibinfo {author}
  {\bibfnamefont {W.}~\bibnamefont {Kozlowski}},\ and\ \bibinfo {author}
  {\bibfnamefont {S.}~\bibnamefont {Wehner}},\ }\href
  {https://doi.org/10.1088/2058-9565/ac753f} {\bibfield  {journal} {\bibinfo
  {journal} {Quantum Science and Technology}\ }\textbf {\bibinfo {volume}
  {7}},\ \bibinfo {pages} {035023} (\bibinfo {year} {2022})}\BibitemShut
  {NoStop}%
\bibitem [{\citenamefont {Maier}(2020)}]{thesis_maier}%
  \BibitemOpen
  \bibfield  {author} {\bibinfo {author} {\bibfnamefont {D.}~\bibnamefont
  {Maier}},\ }\emph {\bibinfo {title} {Investigating the Scalability of Quantum
  Repeater Protocols Based on Atomic Ensembles}},\ \href@noop {} {Master's
  thesis},\ \bibinfo  {school} {Delft University of Technology and Ludwig
  Maximilian University of Munich} (\bibinfo {year} {2020}),\ \bibinfo {note}
  {\href{http://resolver.tudelft.nl/uuid:04b9f054-2139-4b30-ba67-4b7b4752ce86}{http://resolver.tudelft.nl/uuid:04b9f054-2139-4b30-ba67-4b7b4752ce86}}\BibitemShut
  {NoStop}%
\bibitem [{\citenamefont {Labay~Mora}(2021)}]{thesis_Labay_Mora}%
  \BibitemOpen
  \bibfield  {author} {\bibinfo {author} {\bibfnamefont {A.}~\bibnamefont
  {Labay~Mora}},\ }\emph {\bibinfo {title} {Genetic algorithm-based
  optimisation of entanglement distribution to minimise hardware cost}},\
  \href@noop {} {Master's thesis},\ \bibinfo  {school} {Delft University of
  Technology} (\bibinfo {year} {2021}),\ \bibinfo {note}
  {\href{http://resolver.tudelft.nl/uuid:5dd40a56-8c8d-4766-a2fe-0a8c45e1ee3f}{http://resolver.tudelft.nl/uuid:5dd40a56-8c8d-4766-a2fe-0a8c45e1ee3f}}\BibitemShut
  {NoStop}%
\bibitem [{\citenamefont {Zang}\ \emph {et~al.}(2022)\citenamefont {Zang},
  \citenamefont {Kolar}, \citenamefont {Chung}, \citenamefont {Suchara},
  \citenamefont {Zhong},\ and\ \citenamefont
  {Kettimuthu}}]{sequence_absorptive}%
  \BibitemOpen
  \bibfield  {author} {\bibinfo {author} {\bibfnamefont {A.}~\bibnamefont
  {Zang}}, \bibinfo {author} {\bibfnamefont {A.}~\bibnamefont {Kolar}},
  \bibinfo {author} {\bibfnamefont {J.}~\bibnamefont {Chung}}, \bibinfo
  {author} {\bibfnamefont {M.}~\bibnamefont {Suchara}}, \bibinfo {author}
  {\bibfnamefont {T.}~\bibnamefont {Zhong}},\ and\ \bibinfo {author}
  {\bibfnamefont {R.}~\bibnamefont {Kettimuthu}},\ }in\ \href
  {https://doi.org/10.1109/QCE53715.2022.00084} {\emph {\bibinfo {booktitle}
  {2022 IEEE International Conference on Quantum Computing and Engineering
  (QCE)}}}\ (\bibinfo {year} {2022})\ pp.\ \bibinfo {pages}
  {617--623}\BibitemShut {NoStop}%
\bibitem [{\citenamefont {Wu}\ \emph {et~al.}(2021{\natexlab{b}})\citenamefont
  {Wu}, \citenamefont {Kolar}, \citenamefont {Chung}, \citenamefont {Jin},
  \citenamefont {Kettimuthu},\ and\ \citenamefont
  {Suchara}}]{sequence_parallel}%
  \BibitemOpen
  \bibfield  {author} {\bibinfo {author} {\bibfnamefont {X.}~\bibnamefont
  {Wu}}, \bibinfo {author} {\bibfnamefont {A.}~\bibnamefont {Kolar}}, \bibinfo
  {author} {\bibfnamefont {J.}~\bibnamefont {Chung}}, \bibinfo {author}
  {\bibfnamefont {D.}~\bibnamefont {Jin}}, \bibinfo {author} {\bibfnamefont
  {R.}~\bibnamefont {Kettimuthu}},\ and\ \bibinfo {author} {\bibfnamefont
  {M.}~\bibnamefont {Suchara}},\ }\href@noop {} {\bibinfo {title} {Parallel
  simulation of quantum networks with distributed quantum state management}}
  (\bibinfo {year} {2021}{\natexlab{b}}),\ \Eprint
  {https://arxiv.org/abs/2111.03918} {ArXiv:2111.03918 [quant-ph]} \BibitemShut
  {NoStop}%
\bibitem [{\citenamefont {Bennett}\ \emph {et~al.}(2020)\citenamefont
  {Bennett}, \citenamefont {Melchers},\ and\ \citenamefont {Proppe}}]{hpc_fub}%
  \BibitemOpen
  \bibfield  {author} {\bibinfo {author} {\bibfnamefont {L.}~\bibnamefont
  {Bennett}}, \bibinfo {author} {\bibfnamefont {B.}~\bibnamefont {Melchers}},\
  and\ \bibinfo {author} {\bibfnamefont {B.}~\bibnamefont {Proppe}},\
  }\href@noop {} {\bibinfo {title} {{Curta: A general-purpose high-performance
  computer at ZEDAT, Freie Universit\"at Berlin}}},\ \bibinfo {howpublished}
  {DOI:
  \href{https://dx.doi.org/10.17169/refubium-26754}{10.17169/refubium-26754}}
  (\bibinfo {year} {2020})\BibitemShut {NoStop}%
\bibitem [{\citenamefont {Dür}\ and\ \citenamefont
  {Briegel}(2007)}]{epp_qec_review}%
  \BibitemOpen
  \bibfield  {author} {\bibinfo {author} {\bibfnamefont {W.}~\bibnamefont
  {Dür}}\ and\ \bibinfo {author} {\bibfnamefont {H.~J.}\ \bibnamefont
  {Briegel}},\ }\href {https://doi.org/10.1088/0034-4885/70/8/R03} {\bibfield
  {journal} {\bibinfo  {journal} {Rep. Prog. Phys.}\ }\textbf {\bibinfo
  {volume} {70}},\ \bibinfo {pages} {1381} (\bibinfo {year}
  {2007})}\BibitemShut {NoStop}%
\bibitem [{\citenamefont {Lo}\ \emph {et~al.}(2005)\citenamefont {Lo},
  \citenamefont {Chau},\ and\ \citenamefont {Ardehali}}]{Lo2005}%
  \BibitemOpen
  \bibfield  {author} {\bibinfo {author} {\bibfnamefont {H.-K.}\ \bibnamefont
  {Lo}}, \bibinfo {author} {\bibfnamefont {H.~F.}\ \bibnamefont {Chau}},\ and\
  \bibinfo {author} {\bibfnamefont {M.}~\bibnamefont {Ardehali}},\ }\href
  {https://doi.org/10.1007/s00145-004-0142-y} {\bibfield  {journal} {\bibinfo
  {journal} {J. Crypt.}\ }\textbf {\bibinfo {volume} {18}},\ \bibinfo {pages}
  {133} (\bibinfo {year} {2005})}\BibitemShut {NoStop}%
\bibitem [{\citenamefont {Scarani}\ \emph {et~al.}(2009)\citenamefont
  {Scarani}, \citenamefont {Bechmann-Pasquinucci}, \citenamefont {Cerf},
  \citenamefont {Du\ifmmode~\check{s}\else \v{s}\fi{}ek}, \citenamefont
  {L\"utkenhaus},\ and\ \citenamefont {Peev}}]{Scarani2009}%
  \BibitemOpen
  \bibfield  {author} {\bibinfo {author} {\bibfnamefont {V.}~\bibnamefont
  {Scarani}}, \bibinfo {author} {\bibfnamefont {H.}~\bibnamefont
  {Bechmann-Pasquinucci}}, \bibinfo {author} {\bibfnamefont {N.~J.}\
  \bibnamefont {Cerf}}, \bibinfo {author} {\bibfnamefont {M.}~\bibnamefont
  {Du\ifmmode~\check{s}\else \v{s}\fi{}ek}}, \bibinfo {author} {\bibfnamefont
  {N.}~\bibnamefont {L\"utkenhaus}},\ and\ \bibinfo {author} {\bibfnamefont
  {M.}~\bibnamefont {Peev}},\ }\href
  {https://doi.org/10.1103/RevModPhys.81.1301} {\bibfield  {journal} {\bibinfo
  {journal} {Rev. Mod. Phys.}\ }\textbf {\bibinfo {volume} {81}},\ \bibinfo
  {pages} {1301} (\bibinfo {year} {2009})}\BibitemShut {NoStop}%
\bibitem [{\citenamefont {Tomamichel}\ \emph {et~al.}(2012)\citenamefont
  {Tomamichel}, \citenamefont {Lim}, \citenamefont {Gisin},\ and\ \citenamefont
  {Renner}}]{Tomamichel2012}%
  \BibitemOpen
  \bibfield  {author} {\bibinfo {author} {\bibfnamefont {M.}~\bibnamefont
  {Tomamichel}}, \bibinfo {author} {\bibfnamefont {C.~C.~W.}\ \bibnamefont
  {Lim}}, \bibinfo {author} {\bibfnamefont {N.}~\bibnamefont {Gisin}},\ and\
  \bibinfo {author} {\bibfnamefont {R.}~\bibnamefont {Renner}},\ }\href
  {https://doi.org/10.1038/ncomms1631} {\bibfield  {journal} {\bibinfo
  {journal} {Nature Comm.}\ }\textbf {\bibinfo {volume} {3}},\ \bibinfo {pages}
  {634} (\bibinfo {year} {2012})}\BibitemShut {NoStop}%
\bibitem [{\citenamefont {Bacco}\ \emph {et~al.}(2013)\citenamefont {Bacco},
  \citenamefont {Canale}, \citenamefont {Laurenti}, \citenamefont {Vallone},\
  and\ \citenamefont {Villoresi}}]{Bacco2013}%
  \BibitemOpen
  \bibfield  {author} {\bibinfo {author} {\bibfnamefont {D.}~\bibnamefont
  {Bacco}}, \bibinfo {author} {\bibfnamefont {M.}~\bibnamefont {Canale}},
  \bibinfo {author} {\bibfnamefont {N.}~\bibnamefont {Laurenti}}, \bibinfo
  {author} {\bibfnamefont {G.}~\bibnamefont {Vallone}},\ and\ \bibinfo {author}
  {\bibfnamefont {P.}~\bibnamefont {Villoresi}},\ }\href
  {https://doi.org/10.1038/ncomms3363} {\bibfield  {journal} {\bibinfo
  {journal} {Nature Comm.}\ }\textbf {\bibinfo {volume} {4}},\ \bibinfo {pages}
  {2363} (\bibinfo {year} {2013})}\BibitemShut {NoStop}%
\bibitem [{\citenamefont {Lim}\ \emph {et~al.}(2021)\citenamefont {Lim},
  \citenamefont {Xu}, \citenamefont {Pan},\ and\ \citenamefont
  {Ekert}}]{Lim2021_PRL}%
  \BibitemOpen
  \bibfield  {author} {\bibinfo {author} {\bibfnamefont {C.~C.-W.}\
  \bibnamefont {Lim}}, \bibinfo {author} {\bibfnamefont {F.}~\bibnamefont
  {Xu}}, \bibinfo {author} {\bibfnamefont {J.-W.}\ \bibnamefont {Pan}},\ and\
  \bibinfo {author} {\bibfnamefont {A.}~\bibnamefont {Ekert}},\ }\href
  {https://doi.org/10.1103/PhysRevLett.126.100501} {\bibfield  {journal}
  {\bibinfo  {journal} {Phys. Rev. Lett.}\ }\textbf {\bibinfo {volume} {126}},\
  \bibinfo {pages} {100501} (\bibinfo {year} {2021})}\BibitemShut {NoStop}%
\bibitem [{\citenamefont {Bennett}(1992)}]{Bennett1992}%
  \BibitemOpen
  \bibfield  {author} {\bibinfo {author} {\bibfnamefont {C.~H.}\ \bibnamefont
  {Bennett}},\ }\href {https://doi.org/10.1103/PhysRevLett.68.3121} {\bibfield
  {journal} {\bibinfo  {journal} {Phys. Rev. Lett.}\ }\textbf {\bibinfo
  {volume} {68}},\ \bibinfo {pages} {3121} (\bibinfo {year}
  {1992})}\BibitemShut {NoStop}%
\bibitem [{\citenamefont {Wallnöfer}\ \emph {et~al.}(2022)\citenamefont
  {Wallnöfer}, \citenamefont {Hahn}, \citenamefont {Gündoğan}, \citenamefont
  {Sidhu}, \citenamefont {Wiesner}, \citenamefont {Walk}, \citenamefont
  {Eisert},\ and\ \citenamefont {Wolters}}]{Satellites}%
  \BibitemOpen
  \bibfield  {author} {\bibinfo {author} {\bibfnamefont {J.}~\bibnamefont
  {Wallnöfer}}, \bibinfo {author} {\bibfnamefont {F.}~\bibnamefont {Hahn}},
  \bibinfo {author} {\bibfnamefont {M.}~\bibnamefont {Gündoğan}}, \bibinfo
  {author} {\bibfnamefont {J.~S.}\ \bibnamefont {Sidhu}}, \bibinfo {author}
  {\bibfnamefont {F.}~\bibnamefont {Wiesner}}, \bibinfo {author} {\bibfnamefont
  {N.}~\bibnamefont {Walk}}, \bibinfo {author} {\bibfnamefont {J.}~\bibnamefont
  {Eisert}},\ and\ \bibinfo {author} {\bibfnamefont {J.}~\bibnamefont
  {Wolters}},\ }\href {https://doi.org/10.1038/s42005-022-00945-9} {\bibfield
  {journal} {\bibinfo  {journal} {Commun. Phys.}\ }\textbf {\bibinfo {volume}
  {5}},\ \bibinfo {pages} {169} (\bibinfo {year} {2022})}\BibitemShut {NoStop}%
\bibitem [{\citenamefont {Pirandola}(2019)}]{pirandola_onerepeater}%
  \BibitemOpen
  \bibfield  {author} {\bibinfo {author} {\bibfnamefont {S.}~\bibnamefont
  {Pirandola}},\ }\href {https://doi.org/10.1038/s42005-019-0147-3} {\bibfield
  {journal} {\bibinfo  {journal} {Commun. Phys.}\ }\textbf {\bibinfo {volume}
  {2}},\ \bibinfo {pages} {51} (\bibinfo {year} {2019})}\BibitemShut {NoStop}%
\end{thebibliography}%
 
\appendix

\section{Models and evaluation}
\label{sec:detailed_model}

In general, the notation and underlying models used in work are derived from those of Refs. \cite{Luong2016, Trenyi2020, ExtendingQuantumLinks} (the scenarios considered in Appendix \ref{sec:known_setups}), but 
adapted in a way that fits all of them 
in one coherent style.
Here, we comment on a few additional details of our model and strategies.

\subsection{Entanglement purification}
\label{sec:explain_epp}
Entanglement purification protocols are a class of quantum protocols using only \emph{local operations and classical communication} (LOCC) to (potentially probabilistically) transform multiple copies of a noisy entangled state into fewer copies of the same states with a higher fidelity. They are one strategy to deal with noise and imperfections that inevitably arise in any realistic scenario.  

In this work we use the DEJMPS \emph{entanglement purification protocol} (EPP) \cite{dejmps}. It is a probabilistic protocol for purifying $\ket{\Phi^+}$ state vectors, taking two noisy states $\rho_{A_1, B_1}$ and $\rho_{A_2, B_2}$ as input.
On both copies $\sqrt{-iX} \otimes \sqrt{iX}$ is applied, before applying a multilateral CNOT operation $\mathrm{CNOT}^{A_1 \rightarrow A_2} \otimes \mathrm{CNOT}^{B_1 \rightarrow B_2}$ and measuring the second pair in the computational basis. The protocol is considered successful if the measurement outcomes coincide, otherwise it is unsuccessful and the remaining pair has to be discarded.
If the initial fidelity of the input states is sufficiently high, a repeated application of this protocol will result in an increased fidelity for the output pair.

As an example, consider the base case of two input pairs in the same state $\rho$, which is diagonal in the Bell basis with coefficient $\lambda_{ij}$. If the desired state vector $\ket{\Phi^+}$ is affected by local Pauli-diagonal noise it will always be diagonal in the Bell basis. As a side note, an arbitrary bipartite state can be brought to this form by probabilistic application of operations (see, \eg, Ref.\   \cite{epp_qec_review}), however, this is not necessary for the protocol to function as off-diagonal entries do not affect the protocol adversely \cite{dejmps}. 
We write
\begin{equation}
  \begin{aligned}
    \rho = 
    &\lambda_{0,0} \Ketbra{\Phi^+}{\Phi^+} + \lambda_{1,0} \Ketbra{\Phi^-}{\Phi^-} + \\ 
    &\lambda_{0,1} \Ketbra{\Psi^+}{\Psi^+} + \lambda_{1,1} \Ketbra{\Psi^-}{\Psi^-}.
  \end{aligned}
\end{equation}
The effective map after a successful 
entanglement purification step is given by
\begin{equation}
 \begin{aligned}
     \widetilde{\lambda}_{0,0} &= \frac{\lambda_{0,0}^2 + \lambda_{1,1}^2}{N}, \quad &\widetilde{\lambda}_{1,0} = \frac{2 \lambda_{0,0}\lambda_{1,1}}{N} ,\\
     \widetilde{\lambda}_{0,1} &= \frac{\lambda_{0,1}^2 + \lambda_{1,0}^2}{N} ,\quad &\widetilde{\lambda}_{1,1} = \frac{2 \lambda_{0,1}\lambda_{1,0}}{N},
 \end{aligned}
\end{equation}
with $N = (\lambda_{0,0} + \lambda_{1,1})^2 + (\lambda_{0,1} + \lambda_{1,0})^2$ 
as a normalization constant that is also the probability of success.

If the initial fidelity $\lambda_{0,0}$ is sufficiently high, a repeated application of this protocol will result in an amplification of that coefficient. In Ref.~\cite{dejmps}, this repeated application is formulated as a repetition protocol, i.e., one performs the above protocol on many initial copies and uses the output copies of successful purification steps as input for the second step. 

In the scenarios considered in this work, it is common that the input pairs used for the DEJMPS protocol are not identical. If one pair has been established between two parties, it will often have to wait in noisy quantum memories until a second pair can be established and the entanglement purification protocol can be performed. In the context of a quantum repeater, it is also important to stress the necessity of communicating the measurement outcomes, which need to be communicated to both parties performing the protocol in order to decide whether the entanglement purification was successful.

\subsection{Key rates}

The \emph{asymptotic key rate} (per time) is lower bounded by 
\begin{equation}
    r \left[1-h(e_x)-f h(e_z)\right]
    \label{eqn:key_rate}
\end{equation}
\cite{Luong2016, Lo2005, Scarani2009}, where $r$ is the raw rate of bits obtained from measurements at the end stations that have been confirmed to correspond to successful entanglement swapping operations at the repeater stations, $h$ is the binary entropy function and $e_{X(Z)}$ represents the quantum bit error rate of measurements in the $X(Z)$ basis. Furthermore, the key rate can be constrained by an error correction inefficiency $f \geq 1$, with $1$ being to the ideal case.

In this work, we obtain a large sample of long-distance links between the end stations of the repeater chain from our simulation and use the sample mean of $r$ and $e_{X(Z)}$ to calculate an estimate for the asymptotic key rate \eqref{eqn:key_rate}. However, it should be noted that in practice the effects of finite-size effects should be carefully considered for cryptographic applications \cite{Tomamichel2012}, \eg,~some special considerations for satellite-based quantum key distribution (because of the small expected block sizes for current experimental parameters) are discussed in 
Refs.~\cite{Bacco2013,Lim2021_PRL}.

It should be noted that the term key rate is used for multiple related quantities in the literature: Some publications instead focus on the key rate per channel use with the yield $Y$ instead of the raw rate $r$. Furthermore, some use $Y/2$ or $r/2$ to obtain a key rate per mode since the B92 protocol \cite{Bennett1992} requires two modes. 
It is also worth mentioning that depending on how the channel uses are counted the key rate per time may either be directly related to the key rate per channel uses (if the potential number of channel uses in a time interval is counted) or not directly related, e.g., if the actual number of photons sent through a channel in a sequential protocol are counted.

\section{Scaling to more repeater links}
\label{sec:appendix_scaling}

The main texts is restricted to discussing setups of up to 32 repeater links, which is both due to using models that are straightforward to interpret, but also because, as Section \ref{sec:multilink} clearly shows, the range of scenarios that are both viable and interesting while using the same parameter sets are inherently limited. 

\begin{figure}
    \includegraphics[width=\linewidth]{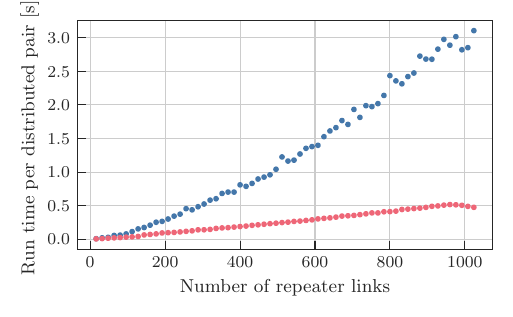}
    \caption{Scaling of run times of the simulation with the number of repeater links. This is for a setup with single memories and no entanglement purification like in Section \ref{sec:multilink} of the main text. Two methods of scheduling events are compared: A general, adjustable method indicative of the protocols used in the rest of this work \textcolor{blue}{(blue)} and a specialized one for this scenario optimized for scaling \textcolor{red}{(red)}. \label{fig:simple_benchmark} }
\end{figure}

In Fig. \ref{fig:simple_benchmark} the run times of repeater setups up to 1024 repeater links are shown. These were obtained on a machine with an Intel Core i9-12900 Processor running up to six data points in parallel.

\section{Known setups with analytical results}
\label{sec:known_setups}

As part of developing the simulation, we naturally tested our approach using known analytic results. In the following, we use the error models and protocols from some specific publications and recreate them using our simulation.
We have done this not only to test whether our simulation works correctly, but also to show that our simulation can support the kind of error models frequently found in the literature. Our framework goes substantially beyond first steps that have been 
taken towards the simulation of repeaters in Ref.\ 
\cite{Satellites}, on which this comprehensive
simulation platform presented here builds.

In the following, we will briefly describe the protocols, explain how the error model translates to our parameter set, and show that our numerically calculated key rates agree with the analytical formulas. All of them use a key rate per channel use as a metric.
 
\subsection{Memory-based quantum repeater with two links}

\begin{figure}
    \centering
    \includegraphics[width=\linewidth]{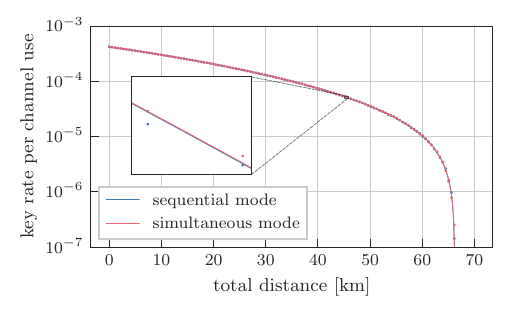}
    \caption{Obtainable key rates for the two link repeater protocol and parameters described in Ref.~\cite{Luong2016}. Comparison of analytic formulas (lines) and our simulation (dots).}
    \label{fig:luetkenhaus}
\end{figure}

 Ref.~\cite{Luong2016} analyses a generic repeater protocol with one repeater station. Entangled pairs are created at the repeater station with one qubit directly loaded into the memory, while the other is sent via glass fibre to the end stations. 
 There are two variants of this protocol, one with sequential entanglement generation, i.e., first only on of the sides tries to establish an entangled pair and only after that is successful, the process is started for the other side, and a simultaneous variant, where both sides are trying to establish pairs simultaneously.

This scenario has an additional parameter for the imperfection that arises from the setups of different repeater stations being misaligned. This can be modeled as a $Y$-noise channel with misalignment error $e_\mathrm{ma}\in [0,1]$ 
acting on one of the qubits
\begin{equation}
    \mathcal{D}^{(i)}_Y (e_\mathrm{ma}) \rho = (1-e_\mathrm{ma}) \rho + e_\mathrm{ma} Y^{(i)} \rho Y^{(i)},
\end{equation}
where $\mathcal{D}_Y$ denotes the Pauli-$Y$-noise channel and $Y$ is the Pauli-$Y$ operator, with the superscript $(i)$ indicating that they each act on the $i$-th qubit.
Note that a $Y$-noise error is consistent with a symmetric, randomly distributed misalignment angle of the detectors for photonic polarization encoded qubits for measurements in the X-Z plane. Both the BB84 and B92 QKD protocols, as well as the Bell state measurements for entanglement swapping, rely on these measurements in the X-Z plane.

 In Fig.~\ref{fig:luetkenhaus}, the results of our simulation compared to the analytical formulas is shown (compare Fig.~3 in Ref.~\cite{Luong2016}). The parameters for this scenario are: 
$P_\mathrm{link} = 0.002376$, 
$T_\mathrm{P} = 2 \times 10^{-6} \text{ s}$,
$e_\mathrm{ma} = 0.01$,
$p_\mathrm{d} = 10^{-8}$,
$\lambda_\mathrm{BSM} = 0.97$,
$F_\mathrm{init} = 1.0$,
$T_\mathrm{dp} = 1.0 \text{ s}$,
$n_\mathrm{mem} = 1$,
$t_\mathrm{cut} = \infty \text{ s}$,
$f = 1.16$.

\medskip

\subsection{Two repeater links for multiple experimental platforms}
\begin{figure*}
    \centering
    \subfloat[]{
        \includegraphics[width=0.48\linewidth]{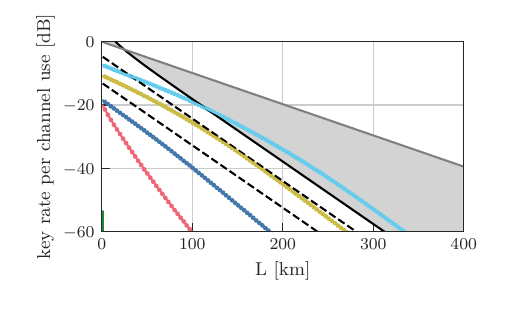}
    } \hfill
    \subfloat[]{
        \includegraphics[width=0.48\linewidth]{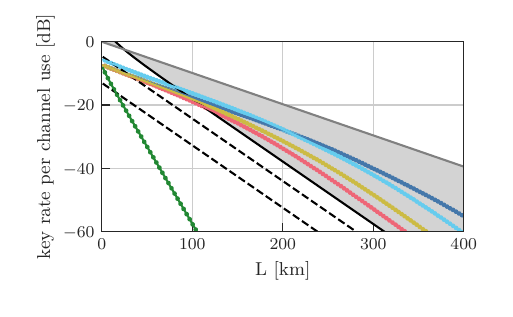}
    }

    \caption{Key rates for node-sends-photons protocols with (a) currently available or (b) future parameters. The experimental platforms are 
    NV-centers (dark blue), SiV-centers (red), Qdots (green), Ca/Yb ions (Yellow) and Rb atoms (light blue), with parameters taken from Ref.~\cite{ExtendingQuantumLinks}. Comparison of 
    analytical formulas (lines) and our simulation (dots).
    The solid black line is the PLOB bound, and the solid grey line equal to $\sqrt{\eta}$ (the asymptotic behavior and upper bound \cite{pirandola_onerepeater} of the ideal one-repeater setup). The two dashed lines are $P_\mathrm{link} \eta / 2$ for $P_\mathrm{link} = 0.7$ and $0.1$. }
    \label{fig:whitepaper}
\end{figure*}

Ref.~\cite{ExtendingQuantumLinks} contains parameters of multiple experimental platforms from research groups in Germany, and summarizes them to a model using only $P_\mathrm{link}$, $T_\mathrm{dp}$ and the clock rate $f_\mathrm{clock}$. The scenario we recreate here is referred to as the \textit{node sends photons} protocol in Ref.~\cite{ExtendingQuantumLinks}, which is a protocol with two repeater links with entangled pair sources located at the central repeater station and also makes use of cut-off times.
The comparison of our simulation results with the analytical formulas is shown in Fig.~\ref{fig:whitepaper} (compare also the secret key plots in Fig.~5 in Ref.~\cite{ExtendingQuantumLinks}). The full parameter sets used are summarized in Tables \ref{tab:whitepaper_current} and \ref{tab:whitepaper_future}.

\begin{table*}[]
    \centering
    \begin{tabular}{|c|c|c|c|c|c|}
        \hline
        & NV & SiV & Qdot & Ca/Yb & Rb \\
        \hline
        $P_\mathrm{link}$ & $0.05$ & $0.05$ & $0.1$ & $0.25$ & $0.5$ \\ \hline
        $T_\mathrm{P}$ & $ 1 / (50 \times 10^6) \text{ s}$ & $ 1 / (30 \times 10^6) \text{ s}$ & $ 1 / (1000 \times 10^6) \text{ s}$ & $ 1 / (0.47 \times 10^6) \text{ s}$ & $ 1 / (5 \times 10^6) \text{ s}$\\ \hline
        $e_\mathrm{ma}$ & $0$ & $0$ & $0$ & $0$ & $0$ \\ \hline
        $p_\mathrm{d}$ & $0$ & $0$ & $0$ & $0$ & $0$ \\ \hline
        $\lambda_\mathrm{BSM}$ & $1.0$ & $1.0$ & $1.0$ & $1.0$ & $1.0$ \\ \hline
        $F_\mathrm{init}$ & $1.0$ & $1.0$ & $1.0$ & $1.0$ & $1.0$ \\ \hline
        $T_\mathrm{dp}$ & $10 \times 10^{-3} \text{ s}$ & $1 \times 10^{-3} \text{ s}$ & $0.003 \times 10^{-3} \text{ s}$ & $20 \times 10^{-3} \text{ s}$ & $100 \times 10^{-3} \text{ s}$ \\ \hline
        $n_\mathrm{mem}$ & $1$ & $1$ & $1$ & $1$ & $1$ \\ \hline
        $t_\mathrm{cut}$ & $ 25 \times L/C$ & $ 10 \times L/C$ & $ 0 \times L/C$ & $ 20 \times L/C$ & $ 100 \times L/C$ \\ \hline
        $f$ & $1.0$ & $1.0$ & $1.0$ & $1.0$ & $1.0$ \\
        \hline
    \end{tabular}
    \caption{Currently available parameters of multiple experimental platforms according to 
    Ref.~\cite{ExtendingQuantumLinks}.}
    \label{tab:whitepaper_current}
\end{table*}

\begin{table*}[]
    \centering
    \begin{tabular}{|c|c|c|c|c|c|}
        \hline
        & NV & SiV & Qdot & Ca/Yb & Rb \\
        \hline
        $P_\mathrm{link}$ & $0.5$ & $0.5$ & $0.6$ & $0.5$ & $0.7$ \\ \hline
        $T_\mathrm{P}$ & $ 1 / (250 \times 10^6) \text{ s}$ & $ 1 / (500 \times 10^6) \text{ s}$ & $ 1 / (1000 \times 10^6) \text{ s}$ & $ 1 / (10 \times 10^6) \text{ s}$ & $ 1 / (10 \times 10^6) \text{ s}$\\ \hline
        $e_\mathrm{ma}$ & $0$ & $0$ & $0$ & $0$ & $0$ \\ \hline
        $p_\mathrm{d}$ & $0$ & $0$ & $0$ & $0$ & $0$ \\ \hline
        $\lambda_\mathrm{BSM}$ & $1.0$ & $1.0$ & $1.0$ & $1.0$ & $1.0$ \\ \hline
        $F_\mathrm{init}$ & $1.0$ & $1.0$ & $1.0$ & $1.0$ & $1.0$ \\ \hline
        
$T_\mathrm{dp}$ & $10000 \times 10^{-3} \text{ s}$ & $100 \times 10^{-3} \text{ s}$ & $0.3 \times 10^{-3} \text{ s}$ & $300 \times 10^{-3} \text{ s}$ & $1000 \times 10^{-3} \text{ s}$ \\ \hline
        $n_\mathrm{mem}$ & $1$ & $1$ & $1$ & $1$ & $1$ \\ \hline
        $t_\mathrm{cut}$ & $ 500 \times L/C$ & $ 50 \times L/C$ & $ 0 \times L/C$ & $ 200 \times L/C$ & $ 500 \times L/C$ \\ \hline
        $f$ & $1.0$ & $1.0$ & $1.0$ & $1.0$ & $1.0$ \\
        \hline
    \end{tabular}
    \caption{Parameters that are potentially achievable in the future for multiple experimental platforms according to Ref.~\cite{ExtendingQuantumLinks}.}
    \label{tab:whitepaper_future}
\end{table*}
 
\subsection{Two repeater links with multiple memories}

Ref.~\cite{Trenyi2020} analyzes a protocol that makes use of multiple memories and obtains expressions for achievable key rates depending on the number of memories. It assumes that there is a separate channel for each quantum memory and trials to establish entangled pairs can be attempted simultaneously. However, unlike the scenarios we discuss in the main text, this is not done for both repeater links at the same time. The protocol works as follows:

First, trials are repeated for one of the repeater links until at least one attempt is successful. It is possible that multiple pairs are established in the same trial. Then, trials are performed for the other repeater link until at least one attempt is successful. The quantum memories now containing qubits from successful attempts are paired and entanglement swapping is performed. Finally, if one of the repeater links had more successes then the other, the leftover qubits are discarded.
This last step has a similar effect as introducing a cut-off time mechanism, as it prevents qubits being stored in quantum memories for too long.

\begin{figure*}
    \centering
    \subfloat[]{
        \includegraphics[width=0.48\linewidth]{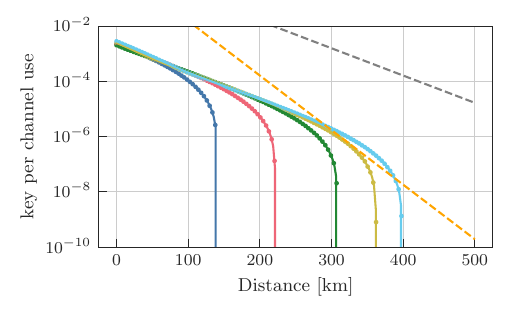}
    } \hfill
    \subfloat[]{
        \includegraphics[width=0.48\linewidth]{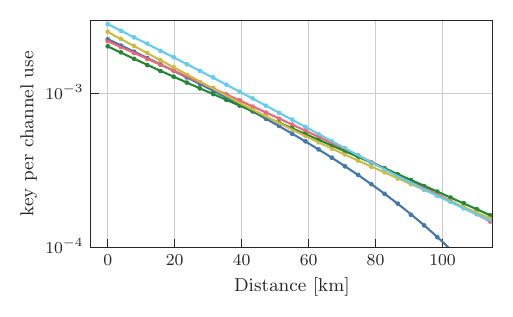}
    }
    \caption{Obtainable key rates for the multi-memory protocol with two repeater links in Ref.~\cite{Trenyi2020}. Comparison of analytical formulas (lines) and our simulation (dots). Results for using 1 (dark blue), 10 (red), 100 (green), 400 (yellow) or 1000 (light blue) memories per repeater link.
    The orange dashed line is the repeater-less PLOB bound, and the gray dashed line is the upper bound for the 1-repeater rate \cite{pirandola_onerepeater}.
    }
    \label{fig:multimemory_luetkenhaus}
\end{figure*}

We present the key rates calculated from our simulation compared to the analytical expressions in 
Fig.~\ref{fig:multimemory_luetkenhaus}. (see also Fig.~2 in 
Ref.~\cite{Trenyi2020}).
The parameters for this scenario are: 
$P_\mathrm{link} = 0.0115$, 
$T_\mathrm{P} = 2 \times 10^{-6} \text{ s}$,
$e_\mathrm{ma} = 0.01$,
$p_\mathrm{d} = 1.8 \times 10^{-11}$,
$\lambda_\mathrm{BSM} = 0.98$,
$F_\mathrm{init} = 1.0$,
$T_\mathrm{dp} = 2.0 \text{ s}$,
$t_\mathrm{cut} = \text{alternative mechanism}$,
$f = 1.16$.

\end{document}